\documentclass[10pt,conference]{IEEEtran} 
\IEEEoverridecommandlockouts

\usepackage{cite}
\usepackage{amsmath,amssymb,amsfonts}
\usepackage{algorithmic}
\usepackage{graphicx}
\usepackage{textcomp}
\usepackage{xcolor}
\usepackage{xspace}
\usepackage{multirow}
\usepackage{booktabs}
\usepackage{arydshln}
\usepackage{tcolorbox}

\usepackage[T1]{fontenc}
\def\BibTeX{{\rm B\kern-.05em{\sc i\kern-.025em b}\kern-.08em
    T\kern-.1667em\lower.7ex\hbox{E}\kern-.125emX}}

\newcommand{\tool}{SECRET\xspace}
\newcommand{\wcgu}[1]{\textcolor{black}{{#1}}}
\newcommand{\revise}[1]{\textcolor{black}{{#1}}}

\newcommand{\secmargin}{\vspace{-1mm}} 
\newcommand{\figmargin}{\vspace{-3mm}} 
\newcommand{\tabmargin}{\vspace{-1.5mm}} 
\newcommand{\eqmargin}{\vspace{-2pt}} 

\usepackage{enumitem}

\begin{document}

\title{SECRET: Towards Scalable and Efficient Code Retrieval via Segmented Deep Hashing}

\author{\IEEEauthorblockN{Wenchao Gu\IEEEauthorrefmark{1}\thanks{\IEEEauthorrefmark{1}Work done while the author was an intern at Microsoft Research.}}
\IEEEauthorblockA{\textit{The Chinese University of Hong Kong}\\
wcgu@cse.cuhk.edu.hk}
\and
\IEEEauthorblockN{Ensheng Shi\IEEEauthorrefmark{1}}
\IEEEauthorblockA{\textit{Xi’an Jiaotong University}\\
s1530129650@stu.xjtu.edu.cn}
\and
\IEEEauthorblockN{Yanlin Wang\IEEEauthorrefmark{2}\IEEEauthorrefmark{3}\thanks{\IEEEauthorrefmark{2}Yanlin Wang is the corresponding author.}\thanks{\IEEEauthorrefmark{3}Work done while the author was a fulltime researcher at Microsoft.}}
\IEEEauthorblockA{\textit{Sun Yat-sen University}\\
yanlin-wang@outlook.com}
\and
\IEEEauthorblockN{Lun Du\IEEEauthorrefmark{3}}
\IEEEauthorblockA{\textit{Ant Research}\\
dulun.dl@antgroup.com}
\and
\IEEEauthorblockN{Shi Han}
\IEEEauthorblockA{\textit{Microsoft}\\
shihan@microsoft.com}
\and
\IEEEauthorblockN{Hongyu Zhang}
\IEEEauthorblockA{\textit{Chongqing University}\\
hyzhang@cqu.edu.cn}
\and
\IEEEauthorblockN{Dongmei Zhang}
\IEEEauthorblockA{\textit{Microsoft}\\
dongmeiz@microsoft.com}
\and
\IEEEauthorblockN{Michael R. Lyu}
\IEEEauthorblockA{\textit{The Chinese University of Hong Kong}\\
lyu@cse.cuhk.edu.hk}
}

\maketitle

\begin{abstract}
\wcgu{Code retrieval, which retrieves code snippets based on users' natural language descriptions, is widely used by developers and plays a pivotal role in real-world software development. The advent of deep learning has shifted the retrieval paradigm from lexical-based matching towards leveraging deep learning models to encode source code and queries into vector representations, facilitating code retrieval according to vector similarity. Despite the effectiveness of these models, managing large-scale code database presents significant challenges. Previous research proposes deep hashing-based methods, which generate hash codes for queries and code snippets and use Hamming distance for rapid recall of code candidates. However, this approach's reliance on linear scanning of the entire code base limits its scalability.} To further improve the efficiency of large-scale code retrieval, we propose a novel approach \tool (Scalable and Efficient Code Retrieval via SegmEnTed deep hashing). \tool converts long hash codes calculated by existing deep hashing approaches into several short hash code segments through an \wcgu{iterative} training strategy. After training, \tool recalls code candidates by looking up the hash tables for each segment, the time complexity of recall can thus be greatly reduced. Extensive experimental results demonstrate that \tool can drastically reduce the retrieval time by at least 95\% while achieving comparable or even higher performance \wcgu{of existing} deep hashing approaches. \wcgu{Besides, \tool also exhibits superior performance and efficiency compared to
the classical hash table-based approach known as LSH under the same number of hash tables.}
\end{abstract}


\secmargin
\section{Introduction}
\secmargin

\wcgu{Code retrieval, a technology enabling the search for relevant code within a codebase using natural language, has garnered significant attention. Due to its pivotal role in real-world software development, many code search approaches~\cite{BrandtGLDK09,McMillanGPXF11,LvZLWZZ15,shi2023cocosoda,SachdevLLKS018,CambroneroLKS019,YaoPS19,ZengYLXWGBDL23} have been proposed in recent years. The open-source communities such as GitHub and Stack Overflow also provide a huge amount of open-source code with natural language descriptions, making it possible to adopt deep learning-based models for code retrieval~\cite{GuZ018,abs-1909-09436,GuLGWZXL21}. Deep learning approaches employing the dual encoder architecture have become predominent in code retrieval tasks~\cite{GuZ018,GuLGWZXL21,CambroneroLKS019}. These approaches use two encoders to encode source code and queries separately into representation vectors. After the encoding, dense retrieval is adopted to retrieve the representation vectors of the source code, which have a strong similarity (e.g., inner product) with the given representation vector of the query~\cite{CambroneroLKS019,ParvezACRC21}.}

\wcgu{However, efficiency in code retrieval for large-scale code databases remains a significant challenge. Since dense retrieval requires a linear scan of the whole code database, the calculation cost of the retrieval for a single query will be extremely high.  The engineering team at GitHub also has underscored the unique challenges posed by GitHub's massive scale. Their search engine encompasses more than 45 million repositories and a vast 115 TB of code. Previous retrieval approaches have struggled with efficiency, resulting in subpar user experiences\footnote{https://github.blog/2023-02-06-the-technology-behind-githubs-new-code-search/}. Thus, effectively searching code within an extremely large code database using natural language descriptions has become a crucial issue in large-scale code retrieval.}

\wcgu{To enhance code retrieval efficiency, Gu et al.~\cite{GuWD0HZL22} introduced a deep hashing-based method called CoSHC. \revise{Deep hashing is a technique which leverages deep learning to convert high-dimensional data into low-dimensional hash codes, facilitating efficient data retrieval and storage. This mapping function preserves the high-dimensional features of the data, ensuring that similar data yield similar hash codes in the hash space~\cite{LuoWWCDHH23,SinghG22,RodriguesCC20}. A typical deep hashing approach uses a deep neural network, such as a CNN or Transformer, similar to those employed for representation learning, to extract features from the original data. After feature extraction, the network maps these features to a low-dimensional hash code space, usually through three fully connected layers and a sign function. The sign function is often not used during the training stage but is adopted during inference to convert the continuous output from the deep learning model into final binary hash code. Unlike methods that rely on cosine similarity between vectors, deep hashing approaches evaluate the similarity between data points by measuring the Hamming distance between their hash codes. CoSHC utilizes a ``recall-rerank'' pipeline, which initially retrieves code candidates based on the Hamming distance of hash codes and then reorders them according to semantic similarity with the query.} By adopting deep hashing, their approach significantly reduces computational costs. While calculating the Hamming distance between binary hash codes can be efficiently performed using the \texttt{XOR} instruction on modern computer architectures~\cite{WangLKC16}, CoSHC still requires scanning the entire large-scale code database for Hamming-distance calculations. Consequently, its retrieval efficiency remains a notable consideration, especially with extremely large code databases.}

To further improve the efficiency of previous deep hashing-based approaches, we propose a code retrieval acceleration framework \tool. By converting the long hash codes from the previous deep hashing approaches into several segmented hash codes and constructing the hash tables for these hash codes, \tool can retrieve the code candidates with lookup hash tables and the retrieval time can be greatly reduced.

Specifically, \tool splits the long hash code from previous deep hashing-based approaches into several segmented hash codes. Hash tables will be constructed for each segmented hash code. During the code retrieval, the segmented hash codes of the given query will be looked up in the corresponding hash table, and the code candidates that are hit in any hash table will be retrieved for the ``re-rank'' step. To reduce the possibility of hash collision between the unmatched codes and queries, \tool proposes the strategy named dynamic matching objective adjustment, which tries to assign a unique hash value for each segmented hash code in the matched pair of code and query. To reduce the difficulty of the hash code alignment between the code and query modality, \tool proposes another strategy named adaptive bit relaxing, which allows \tool to give up the prediction of the hash bits that are found to be hard to align during the training.

Extensive experiments have been conducted to validate the performance of the proposed approach. Experimental results indicate that \tool can reduce at least 95\% of the retrieval time of current deep hashing approaches meanwhile retaining comparable performance or even outperforming previous deep hashing approaches in the recall step. \wcgu{Additionally, we conducted a comparison between \tool and a conventional hash table-based method known as Locality-sensitive hashing (LSH). Based on our experimental findings, \tool demonstrates superior performance and efficiency compared to LSH when the number of hash tables is kept constant.}

This work makes the following key contributions:
\begin{itemize}
\item We propose a novel approach, \tool, to improve the retrieval efficiency of previous deep hashing-based approaches in the task of code retrieval. \tool is the first approach that can convert the long hash code from the deep hashing-based approach into segmented hash codes for the hash table construction. 
\item \tool proposes the dynamic matching objective adjustment strategy, which allows the \tool to dynamically adjust the hash value for each pair of code and query to reduce the false positive hash collision condition.
\item \tool proposes the adaptive bit relaxing strategy, which allows the \tool to give up the prediction of the hash bits that are hard to align during the training.
\item The comprehensive experiments on benchmarks demonstrate that \tool greatly reduces recall computational complexity while keeping advanced performances of previous deep hashing approaches.
\end{itemize}


\begin{figure}[t]
\centering
\includegraphics[width=0.4\textwidth]{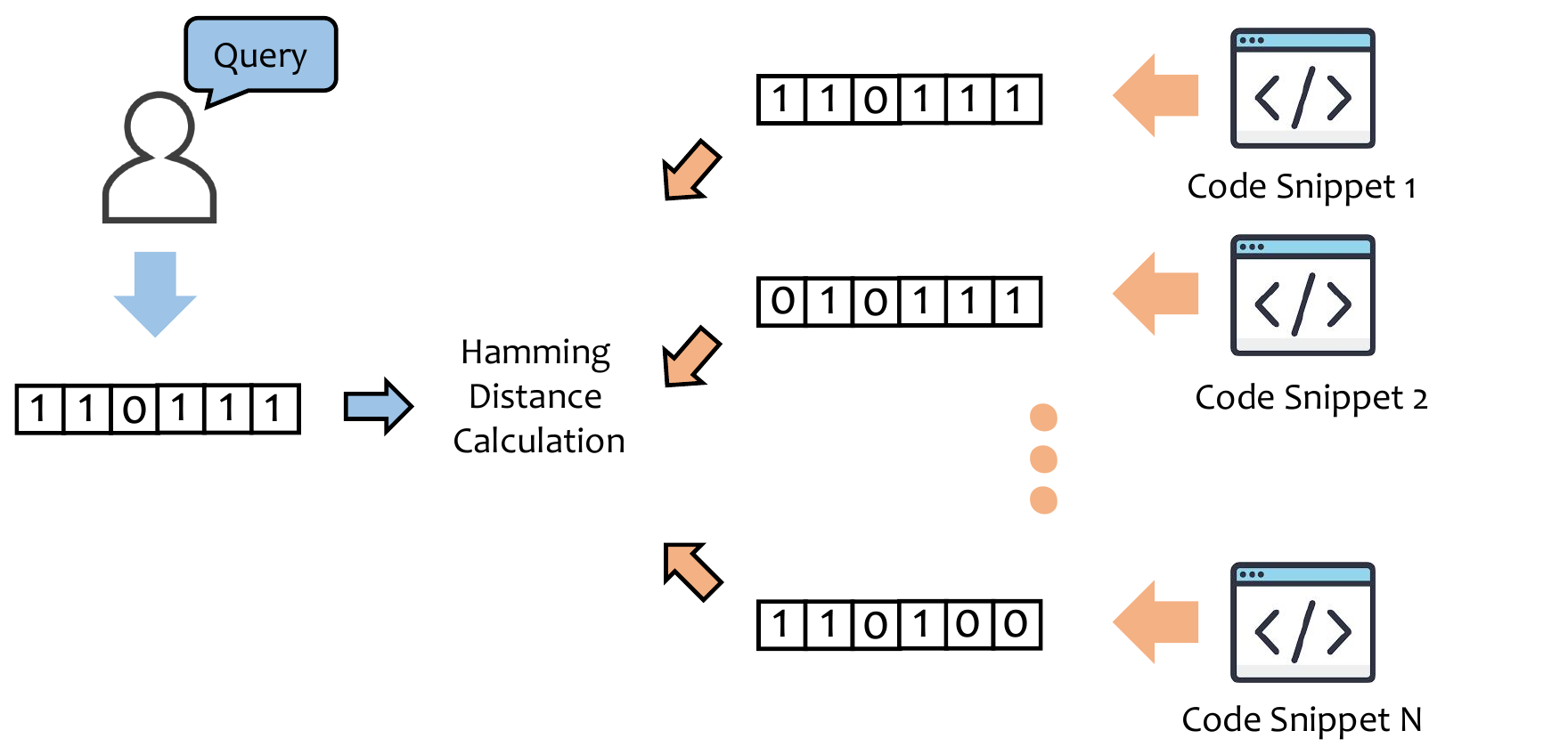}
\caption{\revise{Illustration of recall with previous deep hashing approaches.}}
\label{fig:previous_deep_hashing}
\end{figure}

\begin{figure}[t]
\centering
\includegraphics[width=0.4\textwidth]{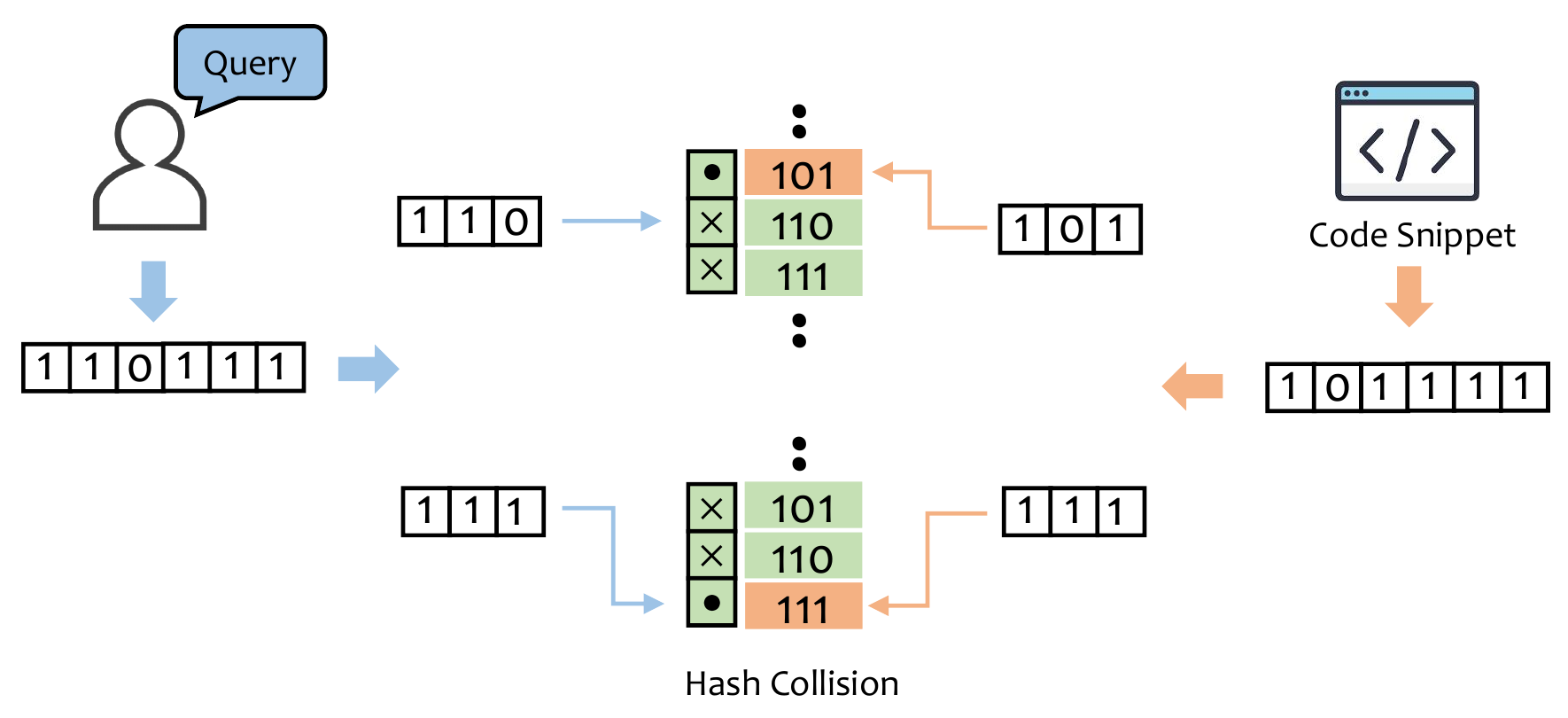}
\caption{\revise{Illustration of recall with the combination of deep hashing approaches and \tool.}}
\label{fig:our_deep_hashing}
\end{figure}

\begin{figure*}[t]
\centering
\includegraphics[width=0.8\textwidth]{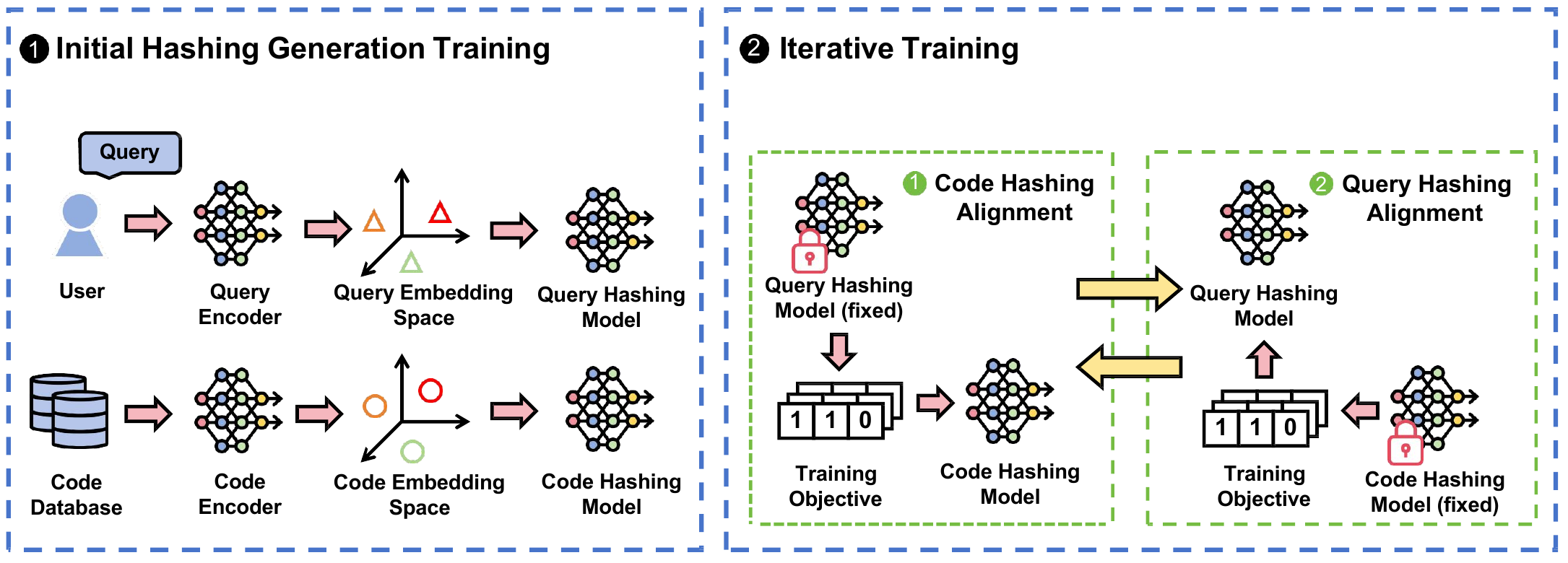}
\figmargin
\caption{Overall framework of \tool. \textit{Initial Hashing Generation Training}:  train the code hashing model and query hashing model with representation vectors from the code encoder and query encoder; \textit{Iterative Training}: consists of two sub-training steps which are code hashing alignment and query hashing alignment. In the sub-training step of hashing alignment, one of the hashing model will be fixed and provide the training objective for training of the other hashing model. These two steps will be performed alternately until the model training converges.
}
\label{fig:framework}
\end{figure*}

\begin{figure*}[t]
\vspace{-5mm}
\centering
\includegraphics[width=0.8\textwidth]{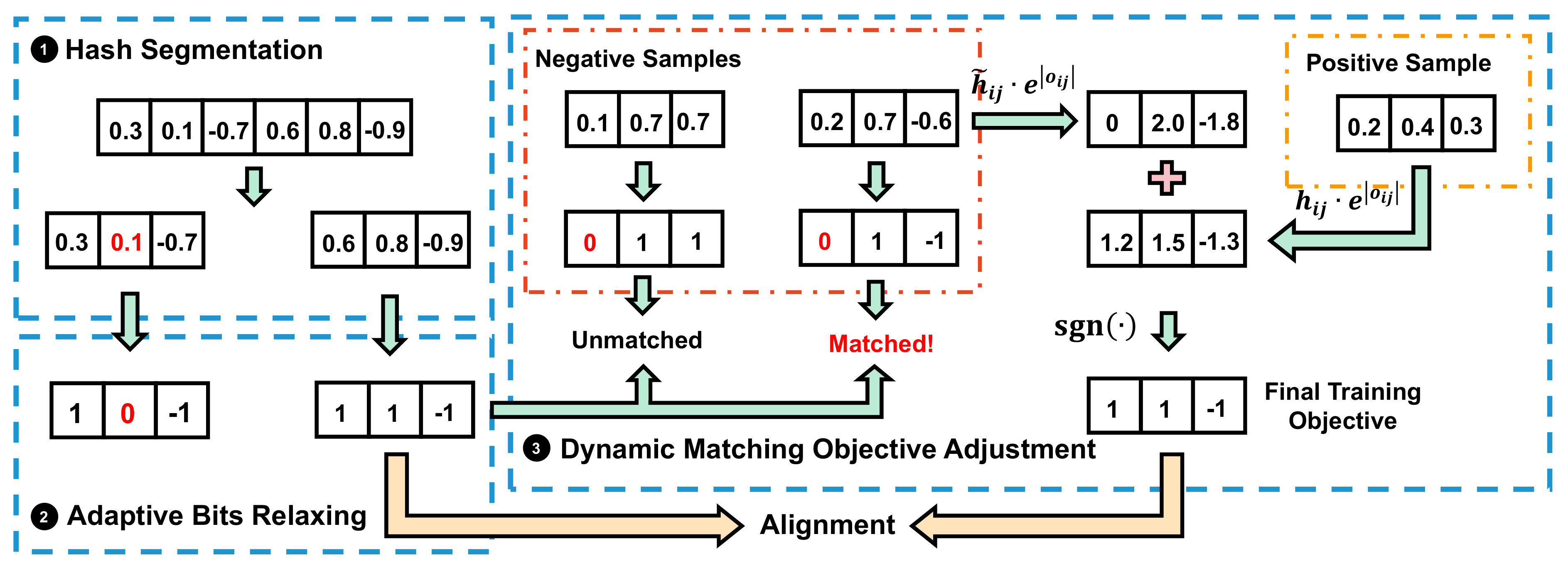}
\figmargin
\caption{Steps in the iterative training strategy. \textit{Hash Segmentation}: split the long hash code into several segmented hash codes and convert continuous output value into the discrete value; \textit{Adaptive Bits Relaxing}: select the hash bits from each segmented hash code according to the absolute output value from the model and overwrite the hash value as ``unknown'' on these hash bits, which represented as 0; \textit{Dynamic Matching Objective Adjustment}: Assign the suitable matching objective for each pair of query and code snippet. The hash code from the positive sample will be utilized as the ground-truth label and adjusted according to the hash collision condition with the negative samples from the same batch.}
\label{fig:algorithm}
\vspace{-5mm}
\end{figure*}

\secmargin
\section{Method}

\secmargin
\subsection{Overview}
\secmargin
\revise{Figure~\ref{fig:previous_deep_hashing} illustrates the recall achieved with deep hashing techniques in code search~\cite{GuWD0HZL22}. In traditional deep hashing approaches, the query and code snippets are encoded into binary hash codes, and code candidates are recalled based on the Hamming distance between the query and code snippet hash codes. In contrast, Figure~\ref{fig:our_deep_hashing} illustrates the recall using a combination of deep hashing techniques and \tool. Unlike previous methods~\cite{GuWD0HZL22}, \tool converts the long hash codes from traditional deep hashing approaches into shorter hash segments. By constructing lookup hash tables for these segments, we can utilize hash collisions instead of Hamming distance calculations during the recall stage, reducing the time complexity from $O(n)$ to $O(1)$.}

\revise{Figure~\ref{fig:framework} depicts the training process for \tool. This process comprises two main stages: the initial hashing generation training stage and the iterative training stage. During the initial hashing generation stage, traditional deep hashing techniques are employed to train two hashing models, which generate initial hash codes for both the code and the query. In the iterative training stage, one hashing model is kept fixed, and its output serves as the training objective for the other hashing model. After training for certain number of epoch, the roles are reversed: the newly trained hashing model becomes fixed, providing the training objective, while the previously fixed hashing model is unlocked for further training. These two steps are repeated iteratively until the model training converges. The goal of this iterative training is to ensure that the output from matched queries and code snippets aligns accurately.}

\secmargin
\subsection{Initial Hashing Generation Training}
\secmargin
We leverage the existing deep hashing approaches to design our initial hashing method for code-query search. In our proposed framework, it contains two individual hashing models for the code modality and query modality, respectively. Each hash model is composed as three fully connection layers with a soft binary transformation module (e.g. Tanh activation). The loss function for the hashing models can be summarized as the following mathematical form:
\begin{equation}
\footnotesize
\eqmargin
\label{equ:original_loss}
    L = \sum_{i}^n f\left(sim(\mathbf{c}^{(i)}, \mathbf{q}^{(i)})\right) + \kappa \cdot \mathbb{E}_{(j, k) \sim P_n} \left[g(sim(\mathbf{c}^{(j)}, \mathbf{q}^{(k)}))\right],
\eqmargin
\end{equation}
where $n$ is the number of positive training pairs, $\mathbf{c}^{(i)}$ is the $i$-th code hashing representation, $\mathbf{q}^{(i)}$ is the $i$-th query hashing representation, $sim(\cdot, \cdot)$ is a similarity function (e.g., cosine similarity or dot production), $f(\cdot)$ is a monotonically decreasing function to rescale the similarity, $g(\cdot)$ a monotonically increasing function correlated to $f(\cdot)$, $P_n$ is a negative sampling distribution from which we can sample a negative pair $(i, j)$, and $\kappa$ is the number of negative samples. After optimization, we will discretize the learned vectors, i.e., if a value is greater than 0, it will be set as 1 otherwise it will be set as 0.

\secmargin
\subsection{Iterative Training}
\secmargin
\wcgu{To construct an effective hash table, it is crucial that the code hashing model and query hashing model produce identical hash codes for matched code and queries. Since deep hashing models produce discrete values as outputs, simply aligning the output of two hashing models may result in the issue of opposite training objectives. To illustrate this problem, consider a scenario where the hash value of the code is 1 and the hash value of the query is -1 for a given pair of code and query. The training objective (target hash value) for modality alignment will differ for the code hashing model and query hashing model, posing challenges for convergence during model training.}

\wcgu{To mitigate this issue, we introduce a novel training strategy called iterative training for hashing model training. During the iterative training stage, one hashing model is trained while the other remains fixed. The output of the fixed hashing model serves as the training objective for the updated hashing model.} With the pre-defined number of training epochs, the fixed model and the updated model will be alternated. The iterative training will be stopped when the hashing models are converged.

There are still two problems remained to solve. Firstly, it is hard to ensure the hash code from the code hashing models and query hashing models to be identical for the matched pair of code and query, which greatly harms the accuracy in the ``recall'' step. Secondly, an ideal hash table should avoid the false positive hash collision, which means the irrelevant code and query should have different hash value. 

\wcgu{To tackle the first problem, we propose two strategies: hash segmentation and adaptive bits relaxation. In the hash segmentation strategy, long hash codes are split into short hash segments for constructing the hash table to reduce the difficulty of hash alignment. The adaptive bits relaxation strategy allows the hashing model to relax hard-to-align hash bits, further easing model alignment challenges.}

\wcgu{Addressing the second problem, we introduce the dynamic matching objective adjustment strategy. Here, the output from the fixed model serves as the temporary training objective and adjusts based on hash collision conditions with hash codes from negative samples in the same training batch. Figure~\ref{fig:algorithm} illustrates the steps in the iterative training strategy, with detailed explanations provided in subsequent sections.}

\subsubsection{Hash Segmentation}\label{sec:hash_segmentation}
\wcgu{To reduce the difficulty of the hash code alignment and increase the hash hit ratio, we split the long hash code from the initial hashing generation into several segments and construct a hash table for each segment. The segmented hash code is}
\begin{equation}
\footnotesize
\eqmargin
    H_i = \{h_{i1},...,h_{ik}\},
\eqmargin
\end{equation}
where $H_i$ is the i-th segmented hash code of the code or query, which is composed of $k$ hash bits from the initial hash code. The j-th hash bit in the i-th segmented hash 
is determined by:
\begin{equation}
\footnotesize
\eqmargin
    h_{ij} = {\rm sgn}(o_{(i-1)*k+j}),
\eqmargin
\end{equation}
where $o_{(i-1)*k+j}$ is the output value of the $((i-1)*k+j)$-th hash bit from the neural network and ${\rm sgn}$ is the sign function. For a more concise representation, $o_{(i-1)*k+j}$ will be replaced by $o_{ij}$ in the following section.

\revise{Let's consider the example illustrated in Figure~\ref{fig:algorithm} (1) for clarification. Assume that the hash code generated by previous deep hashing methods is [0.3, 0.1, -0.7, 0.6, 0.8, -0.9], and the desired length for each hash segment is 3. In this step, the original hash code will be divided into two hash segments: [0.3, 0.1, -0.7] and [0.6, 0.8, -0.9].}

\subsubsection{Adaptive Bits Relaxing}
\label{sec:bits_relaxing}
\wcgu{To further reduce the difficulty of hash alignment, we propose a strategy called \textit{adaptive bits relaxing}. In training, the target output on the hash bit is discrete (+1 and -1). The closer the model's output is to the target value, the better the model's convergence. Consequently, outputs with low absolute values indicate poor convergence and that those hash bits are hard to align. To address these mismatching problems, we omit predictions for these uncertain hash bits and set their outputs to both +1 and -1.}

To achieve the adaptive bits relaxing, we first select the hash bits with top k smallest absolute value as the uncertain bits in each hash segment, which is shown below:
\begin{equation}
\footnotesize
\eqmargin
    S_i = \{j|\left| o_{ij} \right| \rm is \ top \ k \ smallest \ in \ O_i\},
\eqmargin
\end{equation}
where $S_{i}$ is the set that contains the hash bits with the top $k$ smallest absolute value. $k$ is the maximum number of relaxing bits allowed in a single segmented hash code. For these relaxing bits in each hash segment, we replace the initial hash value with 0 as the intermediate value, which is:
\begin{equation}
\footnotesize
\eqmargin
    \tilde{h}_{ij} = \left\{
\begin{aligned}
& 0, \ j \in S_{i} \ {\rm and} \ \left|o_{ij} \right| \leq t\\
& h_{ij}, \quad  \rm otherwise \\
\end{aligned}
\right.,
\eqmargin
\end{equation}
where $\tilde{h}_{ij}$ is the hash code on the i-th hash bit after the relaxing. Since the convergence condition of the model may be good on all the hash bits, we pre-define a threshold value $t$ for the relaxing. Only the hash bits whose absolute value is lower than $t$ are allowed to be relaxed.

\revise{Returning to the example illustrated in Figure ~\ref{fig:algorithm} (2), we proceed from the point where the hash segments have been obtained following the hashing segmentation step. The next step involves relaxing the hash bits. For this example, we assume that the maximum allowed relaxed hash bit is 1 for each hash segment, and the threshold value for relaxation is set at 0.5. The smallest absolute values within each segment are found to be 0.1 and 0.6, respectively. Since only 0.1 is smaller than the predefined threshold value, the second bit in the first hash segment will be relaxed. Consequently, the final hash segments will be [1, 0, -1] and [1, 1, -1].}

\subsubsection{Dynamic Matching Objective Adjustment}
\wcgu{To decrease the false positive ratio during the recall step, we introduce a new strategy called \textit{Dynamic Matching Objective Adjustment}. This method assigns appropriate hash codes to each code-query pair. Initially, we obtain the hash codes from the fixed model as the training objective. Then, we evaluate the hash collision condition of this hash code with negative samples from the training batch. Subsequently, we adjust this training objective based on the hash collision condition with the negative samples and incorporate it into the training process.}

In the first step, we need to check whether the hash code of the negative samples in the batch is the same as the hash code we retrieved from the fixed model. Equation~\ref{equ:check_equ} is the matching results for every bit in the hash code:
\begin{equation}
\footnotesize
\eqmargin
    c_{ij} = \tilde{h}_{ij}^{-} \cdot \tilde{h}_{ij}^{+}
\label{equ:check_equ}
\eqmargin
\end{equation}
\noindent where $\tilde{h}_{ij}^{-}$ is the j-th hash bit in the i-th segmented negative hash codes from the modality which needs to be updated and $\tilde{h}_{ij}^{+}$ is the j-th hash bit in the i-th segmented positive hash codes from the fixed model. $c_{ij}$ indicates whether the j-th hash bit between the i-th segmented positive hash code and the i-th segmented negative hash code is matched. Since we know that $\tilde{h}_{ij}^{-}, \tilde{h}_{ij}^{+} \in \{+1,0,-1\}$, then we can get that $c_{ij} \in \{+1,0,-1\}$.  $c_{ij}= +1$ indicates that the two hash bits are identical, $c_{ij}= 0$ indicates that there is at least one hash bit is relaxed, and $c_{ij}= -1$ indicates that the two hash bits are unmatched. Then we define the $C_{i}$ as follows:
\begin{equation}
\footnotesize
\eqmargin
    C_{i} = {\rm min}\{c_{i1},...,c_{ik}\}
\eqmargin
\end{equation}
\noindent where $C_i = -1$ indicates that there exists at least one hash bit unmatched between two segmented hash codes. otherwise, these two segmented hash codes can be regarded as identical. For the purpose of convenient calculation, we define  $\tilde{C}_i$ to indicate whether the negative segmented hash code has the hash collision with the positive segmented hash code as
\begin{equation}
\footnotesize
\eqmargin
    \tilde{C}_{i} = \left\{
\begin{aligned}
& 0, C_{i} = -1\\
& 1, \rm otherwise \\
\end{aligned}
\right.
\eqmargin
\end{equation}

\noindent where $\tilde{C}_{i} = 1$ indicates that the i-th segmented negative sample has the hash collision with the segmented positive sample, otherwise does not.

Since we have checked the hash collision condition with negative samples, then we can adjust the training objective we get from the fixed model for the hash alignment with such information. The adjusted training objective is determined by: 
\begin{equation}
\footnotesize
\eqmargin
    l_{ij} = h_{ij}^{+} \cdot e^{\gamma \cdot \left|o_{ij}^{+}\right|} - \sum_{n=1}^{m} \tilde{C}_{in} \cdot \tilde{h}_{ij}^{-} \cdot e^{\gamma \cdot \left|o_{ijn}^{-}\right|}
\eqmargin
\end{equation}
\noindent where $l_{ij}$ is the j-th hash bit in the ground-truth label for the i-th segmented hash code. $m$ is the negative sample number in the batch. $\gamma$ is the constant parameter. Since the value range of $o_ij$ is $[\-1,+1]$, $\gamma$ can be utilized to amplify the value range so that there is less probability for the hash bits with good convergence conditions to change their sign. For the positive sample, hash bits before the adaptive bits relaxing are selected in the above equation since we still need to offer a clear optimization target for the neural network and the output of these hash bits may get out of the ill convergence condition in the following training epochs. For the negative samples, the hash bits after the adaptive bits relaxing are selected since the hash collision with the negative samples cannot be avoided by changing the output on these relaxed hash bits. Finally, we need to discrete 
ground-truth label for the hash code bit as:
\begin{equation}
\footnotesize
\eqmargin
    \tilde{l}_{ij} = {\rm sgn}{(l_{ij})}
\eqmargin
\end{equation}

\noindent where $\tilde{l}_{ij}$ is the j-th hash bit in the i-th final segmented hash code template. The final training objective of the i-th segmented hash code for the hash alignment is
\begin{equation}
\footnotesize
\eqmargin
    \tilde{L}_{i} = \{\tilde{l}_{i1},...,\tilde{l}_{ik}\}
\eqmargin
\end{equation}

\revise{Let's revisit the example in Figure~\ref{fig:algorithm} (3). Here, we assume that the hash segments obtained from previous steps are from the query modality (though it doesn't matter whether they are from the query modality or the code modality). We have one positive sample [0.2, 0.4, 0.3] (indicating the corresponding hash segment from the matched code snippets) and two negative samples [0.1, 0.7, 0.7] and [0.2, 0.7, -0.6] (indicating the corresponding hash segments from two randomly selected unmatched code snippets).}

\revise{First, let's focus on the positive sample. By applying the first term in Equation 9, we get the value [1.2, 1.5, -1.3]. Next, we examine the negative samples. The discrete hash value for the hash segment [0.1, 0.7, 0.7] is [0, 1, 1], which does not match the given query hash segment [1, 1, -1], so this negative sample is ignored. The other negative sample matches the given query hash segment. Applying the second term in Equation 9, we get the value [0, 2.0, -1.8]. Finally, we sum these two terms via Equation 9 and apply a sign function to the output. The final training objective is [1, 1, -1], which will be used for training the given query hash segment in subsequent steps.}

\secmargin
\subsection{Hash Alignment}
\secmargin
We align 
hash code with the following cross-entropy loss:
\begin{equation}
\footnotesize
\eqmargin
    L= -(1-\tilde{l}_{ij})*{\rm log}(1-o_{ij}) -(1+\tilde{l}_{ij}) * {\rm log} (1 + o_{ij})
\eqmargin
\end{equation}

\noindent where $\tilde{l}_{ij}$ is the j-th hash bit in the final training objective for the i-th segmented hash code. $o_{ij}$ is the output of j-th hash bit in the i-th segmented hash code from the neural network. 

\secmargin
\subsection{Inference of Binary Hash Codes}
\secmargin
In the inference stage of binary hash codes, binary hash codes of source code and queries will be generated by the corresponding hashing model. Firstly, the hashing model will output the continuous hashing value. Then hash code will be split into several segmented hash codes with adaptive bits relaxing strategy as we introduced in the Subsection~\ref{sec:hash_segmentation} and Subsection~\ref{sec:bits_relaxing}. The unknown state for the hash bit will only be treated as an intermediate state in the inference and finally, it will be converted into both 1 and -1. The hash value of the rest hash bits where be converted into 1 or -1 according to the output hash value as a positive number or a negative number.

During searching with lookup hash tables, all the hit code snippets will be added to the recall candidate set. If the users want to set the maximum size of the recall candidate set, we will use a hash table to count the matched times and then apply a Bucket sort to re-rank these candidates.
\secmargin
\section{Experimental Settings}\label{sec:exp}
\secmargin

\subsection{Research Questions}
\secmargin
\wcgu{In our evaluation, we focus on the following questions:
\begin{itemize}
\item RQ1: What is the Efficiency of \tool?
\item RQ2: What is the Effectiveness of \tool?
\item RQ3: What is the Effectiveness of Adaptive Bits Relaxing and Iterative Training?
\item RQ4: How Many Error Bits Have Been Fixed?
\end{itemize}
}

\wcgu{To investigate RQ1, we conducted efficiency experiments to evaluate the time required in the recall step using \tool compared to previous deep hashing approaches and conventional hashing approach. Due to computational resource limitations, we set the database sizes at 50,000, 100,000, 200,000, and 400,000, respectively. We then analyzed the increasing trend in time for different methods as the data size increased.}

\wcgu{To address RQ2, we conducted performance experiments to compare the performance of \tool with previous deep hashing approaches.}

\wcgu{To investigate RQ3, we tested the performance of model training without the adaptive bit relaxing strategy, as well as directly adopting the adaptive bit relaxation strategy without training, in order to verify their respective contributions to the performance of \tool.}

In RQ4, we evaluated the ratio of mismatched hash bits repaired by the adaptive relaxing strategy and the number of hash bits relaxed by both code and query hashing models. This analysis sheds light on potential resource wastage caused by unnecessary relaxations.

\secmargin
\subsection{Datasets}
\secmargin

\begin{table}[t]
\tabmargin
\centering
\setlength\tabcolsep{12pt}
\caption{Dataset statistics.}
\label{tab:dataset}
\begin{tabular}{llll}
\toprule
\textbf{Dataset} & \textbf{Training} & \textbf{Validation} & \textbf{Test} \\
\midrule
Python & 412,178 & 23,107 & 22,176 \\ 
Java & 454,451 & 15,328 & 26,909  \\ 
\bottomrule
\end{tabular}
\tabmargin
\end{table}

The dataset utilized to evaluate the performance of our proposed approach is provided by CodeBERT~\cite{FengGTDFGS0LJZ20}. CodeBERT selects code snippets and queries from CodeSearchNet~\cite{abs-1909-09436} to construct positive and negative pairs. We remain the positive pairs from the dataset and finally get 412,178 $\rm \left \langle code, query \right \rangle $ pairs as the training set, 23,107 $\rm \left \langle code, query \right \rangle $ pairs as the validation set, and 22,176 $\rm \left \langle code, query \right \rangle $ pairs as the test set in the Python dataset. We get 454,451 $\rm \left \langle code, query \right \rangle $ pairs as the training set, 15,328 $\rm \left \langle code, query \right \rangle $ pairs as the validation set, and 26,909 $\rm \left \langle code, query \right \rangle $ pairs as the test set in the Java dataset, which is shown in Table~\ref{tab:dataset}. \revise{We tested 10,000 queries across databases of varying sizes (50,000, 100,000, 200,000, and 400,000) for research question 1. For the remaining research questions, we tested 22,176 queries using a database of 22,176 Python code snippets and 26,909 queries using a database of 26,909 Java code snippets.}

\secmargin
\subsection{Baselines}
\secmargin
We select two state-of-the-art deep learning-based code retrieval models with four deep hashing approaches and a non-learning-based approach as our baselines.

\subsubsection{Code Retrieval Baselines}
\label{sec:code_baselines}
We select CodeBERT~\cite{FengGTDFGS0LJZ20} and GraphCodeBERT~\cite{GuoRLFT0ZDSFTDC21} as our base code retrieval models. Both are state-of-the-art models in the code retrieval task.

\begin{itemize}[leftmargin=15pt]
\item \textbf{CodeBERT} is a bi-modal pre-trained model based on a Transformer with 12 layers for programming languages and natural languages. 
\item \textbf{GraphCodeBERT} is another pre-trained Transformer-based model. Unlike previous pre-trained models which only utilize the sequence of code tokens as the features, GraphCodeBERT additionally considers the data flow of code snippets in the pre-training stage.
\end{itemize}

\subsubsection{Deep Hashing Baselines} 
We select four state-of-the-art baseline models of deep hashing, which are CoSHC~\cite{GuWD0HZL22}, DJSRH~\cite{SuZZ19}, DSAH~\cite{YangWZZL020}, and JDSH~\cite{LiuQGZY20}. We also select a hash table based approach LSH~\cite{DatarIIM04} and \revise{two non hash table based approaches which are BM25~\cite{RobertsonW94} and TF-IDF~\cite{Jones04}}. 
\begin{itemize}[leftmargin=15pt]
\item \textbf{CoSHC} is the first approach that combines the deep hashing techniques with 
code classification to accelerate the code search. For the sake of fairness of the experiment, we only adopt the deep hashing parts from this approach.
\item \textbf{DJRSH} constructs a novel joint-semantic affinity matrix that contains specific similarity values instead of similarity order as in previous approaches. 
\item \textbf{DSAH} designs a semantic-alignment loss 
to align 
similarity between input features and generated binary hash. 
\item \textbf{JDSH} is a deep hashing approach that jointly trains 
different modalities with a joint-modal similarity matrix, which can fully preserve cross-modal semantic correlations. 
\item \textbf{LSH} is one of the most popular approaches that map high dimensional data to hash value by using random hash functions and constructing 
lookup hash tables for 
data searching. It is widely applied in data recall for 
single modality.
\revise{\item \textbf{BM25} is a widely popular approach for ranking documents, renowned for its effectiveness in search engines and information retrieval systems. It calculates a relevance score by considering the frequency of query terms in a document, the length of the document, and the term's importance across the entire corpus.}
\revise{\item \textbf{TF-IDF} is a popular method in text analysis and information retrieval. It emphasizes terms that are important within a specific document but rare across the entire dataset, making it an effective tool for retrieving relevant data.}
\end{itemize}

\secmargin
\subsection{Metrics}
\secmargin
\revise{We use R@k (recall at $k$), MRR (mean reciprocal rank), and N@k (Normalized discounted cumulative gain at k) as the evaluation metrics in this paper.}
R@k is the metric which widely used to evaluate the performance of the code retrieval models by many previous  studies~\cite{HaldarWXH20,ShuaiX0Y0L20,FangTZL21,abs-2008-12193}. It is defined as:
\begin{equation}
\footnotesize
\vspace{-2pt}
    R@k = \frac{1}{|Q|}\sum^Q_{q=1}\delta(FRank_q \leq k),
\end{equation}
\noindent where $Q$ denotes the query set and $FRank_q$ is the rank of the correct answer for query $q$. $\delta(Frank_q \leq k)$ returns 1 if the correct result is within the top $k$ returning results, otherwise it returns 0. A higher R@k indicates better performance.

MRR is another metric widely used in the code retrieval task to evaluate the performance~\cite{FengGTDFGS0LJZ20,GuoRLFT0ZDSFTDC21}:
\begin{equation}
\footnotesize
\eqmargin
    MRR = \frac{1}{|Q|}\sum^Q_{q=1}\frac{1}{FRank_q}
\eqmargin
\end{equation}
A higher MRR indicates better performance.

\revise{N@k is a metric used to assess the effectiveness of recommendation systems by evaluating both the relevance and the ranking of the results they provide. It is defined as follows:}
\begin{equation}
\footnotesize
\eqmargin
    DCG@k = \sum^k_{i=1}\frac{rel_i}{{\rm log}(i+1)}
\eqmargin
\end{equation}
\begin{equation}
\footnotesize
\eqmargin
IDCG@k = \sum^k_{i=1}\frac{rel^{ideal}_i}{{\rm log}(i+1)}
\eqmargin
\end{equation}
\begin{equation}
\footnotesize
\eqmargin
NDCG@k = \frac{DCG@k}{IDCG@k}
\eqmargin
\end{equation}

\noindent \revise{where $rel_i$ is the relevance score of the retrieved i-th code snippet, and $rel^{ideal}_i$ is the ideal relevance score of the i-th code snippet. We assign a relevance score of 1 to the matched code snippet and 0 to the unmatched code snippets. A higher N@k indicates better performance.}

\begin{table*}[t]
\tabmargin
\scriptsize
\setlength\tabcolsep{7pt}
\revise{\caption{Results of time efficiency comparison on the recall step of different deep hashing approaches with different code retrieval models on the Python dataset with the size 50,000, 100,000, 200,000 and 400,000.}}
\begin{tabular}{llllllllll}
\toprule
& & \multicolumn{2}{c}{\textbf{50,000}} & \multicolumn{2}{c}{\textbf{100,000}} & \multicolumn{2}{c}{\textbf{200,000}} & \multicolumn{2}{c}{\textbf{400,000}}\\

\cmidrule(lr){3-4} \cmidrule(lr){5-6} \cmidrule(lr){7-8}\cmidrule(lr){9-10}
& & \textbf{128bit} & \textbf{256bit} & \textbf{128bit} & \textbf{256bit} & \textbf{128bit} & \textbf{256bit}& \textbf{128bit} & \textbf{256bit}\\
\midrule
\parbox[t]{3mm}{\multirow{9}{*}{\rotatebox[origin=c]{90}{CodeBERT}}} & LSH & 3.8s & 7.5s & 8.1s & 16.4s & 16.7s & 37.6s & 38.8s & 82.8s\\ 
& BM25 & 308.6s & 308.6s & 623.2s & 623.2s & 1258.9s &  1258.9s & 2555.5s & 2555.5s \\
& TF-IDF & 309.2s & 309.2s & 670.9s & 670.9s & 1455.9s & 1455.9s & 3173.8s & 3173.8s \\ 
\cmidrule(lr){2-10}
& CoSHC & 31.9s & 43.7s & 66.4s & 90.1s & 137.7s & 184.8s & 280.1s & 375.7s\\ 
& CoSHC\textsubscript{\tool}  & 1.2s ($\downarrow$96.2\%) & 2.2s ($\downarrow$95.0\%)  & 2.1s ($\downarrow$96.8\%) & 3.8s ($\downarrow$95.8\%) & 4.0s ($\downarrow$97.1\%)  & 7.1s ($\downarrow$96.2\%) & 7.8s ($\downarrow$97.2\%) & 14.2s ($\downarrow$96.2\%) \\ 
\cmidrule(lr){2-10}
& DJSRH & 31.2s & 43.1s & 65.2s & 88.7s & 135.1s & 185.8s & 274.5s & 367.8s\\ 
& DJSRH\textsubscript{\tool}  & 1.2s ($\downarrow$96.2\%) & 1.4s ($\downarrow$96.8\%)  & 2.1s ($\downarrow$96.8\%) & 2.5s ($\downarrow$97.1\%) & 3.9s ($\downarrow$97.1\%)  & 4.4s ($\downarrow$97.6\%) & 7.9s ($\downarrow$97.1\%) & 8.4s ($\downarrow$97.6\%) \\ 
\cmidrule(lr){2-10}
& DSAH & 31.2s & 44.0s  & 65.3s & 90.5s & 135.0s & 186.0s & 275.5s & 376.8s \\
& DSAH\textsubscript{\tool} & 1.0s ($\downarrow$96.8\%) & 1.4s ($\downarrow$96.8\%) & 1.9s ($\downarrow$97.1\%) & 2.5s ($\downarrow$97.2\%) & 3.5s ($\downarrow$97.4\%) & 4.5s ($\downarrow$97.6\%) & 6.8s  ($\downarrow$97.5\%) & 8.4s ($\downarrow$97.8\%)\\
\cmidrule(lr){2-10}
& JDSH & 31.1s & 42.2s & 65.2s & 90.6s & 135.0s & 185.7s
 & 274.4s & 368.6s\\			
& JDSH\textsubscript{\tool} & 1.2s ($\downarrow$96.1\%) & 1.6s ($\downarrow$96.2\%) & 2.0s ($\downarrow$96.9\%) & 2.6s ($\downarrow$97.1\%) & 3.8s ($\downarrow$97.2\%) & 4.6s ($\downarrow$97.5\%) & 7.5s ($\downarrow$97.3\%) & 8.6s ($\downarrow$97.7\%)\\
\midrule
\parbox[t]{3mm}{\multirow{9}{*}{\rotatebox[origin=c]{90}{GraphCodeBERT}}} & LSH & 3.7s & 7.3s & 7.7s & 15.5s & 16.9s & 34.9s & 38.2s & 82.2s \\ 
& BM25 & 308.1s & 308.1s & 622.8s & 622.8s & 1256.1s &  1256.1s & 2552.9s & 2552.9s \\
& TF-IDF & 307.5s & 307.5s & 668.1s & 668.1s & 1450.2s & 1450.2s & 3165.2s & 3165.2s \\
\cmidrule(lr){2-10}
& CoSHC & 31.9s & 43.7s & 66.5s & 90.1s & 137.6s & 184.9s & 280.0s & 375.9s\\ 
& CoSHC\textsubscript{\tool} & 1.1s ($\downarrow$96.6\%) & 2.2s ($\downarrow$95.0\%)  & 2.0s ($\downarrow$97.0\%) & 3.8s ($\downarrow$95.8\%) & 3.8s ($\downarrow$97.2\%)  & 7.0s ($\downarrow$96.2\%) & 7.4s ($\downarrow$97.4\%) & 13.8s ($\downarrow$96.3\%) \\ 
\cmidrule(lr){2-10}
 & DJSRH & 31.2s & 43.0s & 65.2s & 88.5s & 134.9s & 181.7s & 274.5s & 367.8s \\ 
& DJSRH\textsubscript{\tool} & 1.1s ($\downarrow$96.5\%) & 1.5s ($\downarrow$96.5\%)  & 2.0s ($\downarrow$96.9\%) & 2.6s ($\downarrow$97.1\%)  & 3.8s ($\downarrow$97.2\%) & 4.6s ($\downarrow$97.5\%)  & 7.6s ($\downarrow$97.2\%) & 8.8s ($\downarrow$97.6\%) \\ 
\cmidrule(lr){2-10}
& DSAH & 31.1s & 43.9s & 65.2s & 90.4s & 134.9s & 185.7s & 274.7s & 377.4s\\
& DSAH\textsubscript{\tool} & 1.0s ($\downarrow$96.7\%) & 1.5s ($\downarrow$96.6\%)  &  1.8s ($\downarrow$97.2\%) & 2.6s ($\downarrow$97.1\%) & 3.4s ($\downarrow$97.5\%) & 4.7s ($\downarrow$97.5\%) & 6.6s ($\downarrow$97.6\%) & 8.8s ($\downarrow$97.7\%)\\
\cmidrule(lr){2-10}
& JDSH & 31.1s & 42.3s & 65.1s & 90.3s & 134.9s & 185.6s & 275.2s & 355.0s\\
& JDSH\textsubscript{\tool} & 1.1s ($\downarrow$96.5\%) & 1.7s ($\downarrow$96.0\%) &  2.0s($\downarrow$96.9\%) & 2.6s ($\downarrow$97.1\%) & 3.7s ($\downarrow$97.3\%) & 4.9s ($\downarrow$97.3\%) &  7.2s ($\downarrow$97.4\%) & 9.2s ($\downarrow$97.4\%)\\
\bottomrule
\end{tabular}
\tabmargin
\vspace{-5mm}
\label{tab:efficiency_python}
\end{table*}

\secmargin
\subsection{Implementation Details}
\secmargin
We use the dual encoder paradigm to use two CodeBERT or GraphCodeBERT to encode the source codes and queries into representation vectors. The dimension of the representation vectors is 768. 
We implement CoSHC, DJSRH, DSAH, and JDSH by ourselves and follow the hyperparameter settings of deep hashing baselines described their original papers. We experiment on 128-bit and 256-bit for the generated binary hash codes. The hidden size of all deep hashing models is 1,536. We set the size of the binary hash code segment to 16 bits and we allow the deep hash model to predict no more than three unknown bits in each segment. Due to the low recall ratio of LSH, we reduce the length of hash segment into 8 bits. In addition, we set threshold value $t$ as 0.5. In the overall performance comparison experiment in Section~\ref{sec:rq2}, deep hashing models retrieve the top 300 candidates of code snippets at first. Then the code retrieval models 
determine the final ranking order of these candidates. In the time efficiency experiment in Section~\ref{sec:rq1}, we only compare the time cost for the top 300 candidates retrieved by the deep hashing models since our focus is  the retrieval efficiency of the recall step.
The learning rate of the code retrieval models including CodeBERT and GraphCodeBERT is $1e^{-5}$ and the learning rate for all the deep hashing baselines is $1e^{-4}$. All  models are optimized via the AdamW algorithm~\cite{KingmaB14}.

We train our models on a server with Tesla V100. We train CodeBERT or GraphCodeBERT for 10 epochs. The training epoch both either initial hashing projection and iterative training is 100. Early stopping 
is adopted to avoid over-fitting. We evaluate the retrieval efficiency of \tool 
on a server with Intel Xeon E5-2698v4 2.2Ghz 20-core. The code for efficiency evaluation is written in C++ and the program is only allowed to use a single thread of CPU for fair comparison.

\secmargin
\section{Evaluation}
\secmargin

\subsection{RQ1: What is the Efficiency of \tool?}
\label{sec:rq1}
\secmargin

\begin{table*}[t]
\tabmargin
\centering
\setlength\tabcolsep{2pt}
\revise{\caption{Overall performance comparison of different deep hashing approaches with different code retrieval models.}}

\fontsize{5.8}{7}\selectfont
\begin{tabular}{llllllllllllll}
\toprule
 \multirow{3}{*}{\textbf{}}&\multirow{3}{*}{\textbf{Model}} & \multicolumn{6}{c}{\textbf{Python}} & \multicolumn{6}{c}{\textbf{Java}} \\

\cmidrule(lr){3-8} \cmidrule(lr){9-14}
& & \multicolumn{3}{c}{\textbf{128bit}} & \multicolumn{3}{c}{\textbf{256bit}} & \multicolumn{3}{c}{\textbf{128bit}} & \multicolumn{3}{c}{\textbf{256bit}}\\

\cmidrule(lr){3-5} \cmidrule(lr){6-8} \cmidrule(lr){9-11}\cmidrule(lr){12-14}
& & \textbf{R@1} & \textbf{MRR} & \textbf{N@10} & \textbf{R@1} & \textbf{MRR} & \textbf{N@10} & \textbf{R@1} & \textbf{MRR}& \textbf{N@10} & \textbf{R@1} & \textbf{MRR} & \textbf{N@10} \\
 
\midrule
\parbox[t]{4mm}{\multirow{10}{*}{\rotatebox[origin=c]{90}{CodeBERT}}} & Original & 0.455 & 0.562 & 0.606 & 0.455 & 0.563 & 0.606 & 0.322 & 0.420 & 0.459 & 0.322 & 0.420 & 0.459 \\ 
& LSH & 0.388 & 0.461 & 0.491 & 0.441 & 0.533 & 0.572 & 0.265 & 0.331 & 0.358 & 0.303 & 0.387 & 0.422 \\ 
& BM25 & 0.448 & 0.541 & 0.580 & 0.448 & 0.541 & 0.580 & 0.245 & 0.305 & 0.329 & 0.245 & 0.305 & 0.329 \\
& TF-IDF & 0.451 & 0.548 & 0.606 & 0.451 & 0.548 & 0.606 & 0.263 & 0.330 & 0.372 & 0.263 & 0.330 & 0.372 \\
\cmidrule(lr){2-14}
& CoSHC & 0.455 & 0.562 & 0.605 & 0.455 & 0.563 & 0.606 & 0.321 & 0.419 & 0.457 & 0.322 & 0.420 & 0.459 \\ 
& CoSHC\textsubscript{\tool}  & 0.447 ($\downarrow$1.8\%) & 0.547 ($\downarrow$2.7\%) & 0.589 ($\downarrow$2.6\%) & 0.452 ($\downarrow$0.7\%) & 0.554 ($\downarrow$1.6\%) & 0.596 ($\downarrow$1.7\%) & 0.316 ($\downarrow$1.6\%)  & 0.408 ($\downarrow$2.6\%) & 0.445 ($\downarrow$2.6\%) & 0.319 ($\downarrow$0.9\%) & 0.415 ($\downarrow$1.2\%) & 0.454 ($\downarrow$1.1\%) \\ 
\cmidrule(lr){2-14}
 & DJSRH & 0.454 & 0.561 & 0.604 & 0.455 & 0.563 & 0.606 & 0.321 & 0.418 & 0.457 & 0.322 & 0.420 & 0.459 \\ 
& DJSRH\textsubscript{\tool} & 0.446 ($\downarrow$1.8\%) & 0.546 ($\downarrow$2.7\%) 
& 0.587 ($\downarrow$2.8\%)
& 0.451 ($\downarrow$0.9\%) & 0.553 ($\downarrow$1.8\%) 
& 0.596 ($\downarrow$1.7\%)
& 0.316 ($\downarrow$1.6\%)  & 0.409 ($\downarrow$2.2\%) 
& 0.446 ($\downarrow$2.4\%)
& 0.319 ($\downarrow$0.9\%) & 0.414 ($\downarrow$1.4\%) & 0.452 ($\downarrow$1.5\%) \\ 
\cmidrule(lr){2-14}
& DSAH & 0.450 & 0.552 & 0.594  & 0.451 & 0.554 & 0.596 & 0.317 & 0.411 & 0.448 & 0.319 & 0.414 & 0.452\\
& DSAH\textsubscript{\tool} & 0.447 ($\downarrow$0.7\%) & 0.547 ($\downarrow$0.9\%) 
& 0.587 ($\downarrow$1.2\%) 
& 0.452 ($\uparrow$0.2\%) & 0.554 (0.0\%) & 0.596 (0.0\%) & 0.316 ($\downarrow$0.3\%) & 0.409 ($\downarrow$0.5\%) & 0.446 ($\downarrow$0.4\%) & 0.319 (0.0\%) & 0.415 ($\uparrow$0.2\%) & 0.454 ($\uparrow$0.4\%) \\
\cmidrule(lr){2-14}
& JDSH & 0.448 & 0.549 & 0.590 & 0.450 & 0.552 & 0.594 & 0.317 & 0.410 & 0.448 & 0.318 & 0.412 & 0.450\\
& JDSH\textsubscript{\tool} & 0.447 ($\downarrow$0.2\%) & 0.547 ($\downarrow$0.4\%) & 0.587 ($\downarrow$0.5\%)
& 0.452 ($\uparrow$0.4\%) & 0.554 ($\uparrow$0.4\%) & 0.596 ($\uparrow$0.3\%) & 0.316 ($\downarrow$0.3\%) & 0.409 ($\downarrow$0.2\%) & 0.447 ($\downarrow$0.2\%) & 0.319 ($\uparrow$0.3\%) & 0.415 ($\uparrow$0.7\%) & 0.452 ($\uparrow$0.4\%) \\
\midrule
\parbox[t]{4mm}{\multirow{10}{*}{\rotatebox[origin=c]{90}{GraphCodeBERT}}} & Original & 0.489 & 0.598 & 0.641 & 0.489 & 0.598 & 0.641 & 0.355 & 0.457 & 0.498 & 0.355 & 0.457 & 0.498 \\ 
& LSH & 0.409 & 0.478 & 0.506 & 0.469 & 0.562 & 0.600 & 0.281 & 0.343 & 0.368 & 0.330 & 0.415 & 0.450 \\ 
& BM25 & 0.472 & 0.562 & 0.599 & 0.472 & 0.562 & 0.599 & 0.262 & 0.319  & 0.343 & 0.262 & 0.319 & 0.343 \\
& TF-IDF & 0.477 & 0.570 & 0.624 &  0.477 & 0.570 & 0.624  & 0.284 & 0.348 & 0.386 & 0.284 & 0.348 & 0.386  \\
\cmidrule(lr){2-14}
 & CoSHC & 0.489 & 0.597 & 0.640 & 0.489 & 0.598 & 0.641 & 0.355 & 0.455 & 0.496 & 0.355 & 0.457 & 0.498 \\ 
& CoSHC\textsubscript{\tool} & 0.479 ($\downarrow$2.0\%) & 0.580 ($\downarrow$2.8\%) 
& 0.620 ($\downarrow$3.1\%) 
& 0.484 ($\downarrow$1.0\%) & 0.587 ($\downarrow$1.8\%) 
& 0.628 ($\downarrow$2.0\%)
& 0.348 ($\downarrow$2.0\%) & 0.443 ($\downarrow$2.6\%)
& 0.482 ($\downarrow$2.8\%)
& 0.353 ($\downarrow$0.6\%) & 0.451 ($\downarrow$1.3\%) & 0.491 ($\downarrow$1.4\%) \\ 
\cmidrule(lr){2-14}
& DJSRH & 0.489 & 0.597 & 0.640 & 0.489 & 0.598 & 0.641 & 0.354 & 0.454 & 0.495 & 0.355 & 0.457 & 0.498 \\ 
& DJSRH\textsubscript{\tool} & 0.479 ($\downarrow$2.0\%) & 0.579 ($\downarrow$3.0\%)  & 0.619 ($\downarrow$3.3\%)
& 0.482 ($\downarrow$1.4\%) & 0.586 ($\downarrow$2.0\%)  & 0.629 ($\downarrow$1.9\%) 
& 0.348 ($\downarrow$1.7\%) & 0.444 ($\downarrow$2.2\%)  & 0.482 ($\downarrow$2.6\%)
& 0.353 ($\downarrow$0.6\%) & 0.450 ($\downarrow$1.5\%) & 0.490 ($\downarrow$1.6\%) \\ 
\cmidrule(lr){2-14}
& DSAH & 0.482 & 0.586 & 0.628 & 0.484 & 0.589 & 0.631 & 0.351 & 0.447 & 0.486 & 0.352 & 0.449 & 0.488 \\
& DSAH\textsubscript{\tool} & 0.480 ($\downarrow$0.4\%) & 0.580 ($\downarrow$1.0\%)  & 0.620 ($\downarrow$1.3\%) & 0.484 (0.0\%) & 0.587 ($\downarrow$0.3\%) & 0.629 ($\downarrow$0.3\%) & 0.349 ($\downarrow$0.6\%) & 0.444 ($\downarrow$0.7\%) & 0.482 ($\downarrow$0.8\%) & 0.353 ($\uparrow$0.3\%) & 0.450 ($\uparrow$0.2\%) & 0.490 ($\uparrow$0.4\%) \\
\cmidrule(lr){2-14}
& JDSH & 0.482 & 0.585 & 0.629 & 0.483 & 0.587 & 0.629 & 0.350 & 0.446 & 0.485 & 0.351 & 0.448 & 0.487 \\
& JDSH\textsubscript{\tool} & 0.478 ($\downarrow$0.8\%) & 0.579 ($\downarrow$1.0\%) & 0.619 ($\downarrow$1.6\%) & 0.483 (0.0\%) & 0.586 ($\downarrow$0.2\%) & 0.628 ($\downarrow$0.2\%) & 0.349 ($\downarrow$0.3\%) & 0.443 ($\downarrow$0.7\%) & 0.483 ($\downarrow$0.4\%) & 0.353 ($\uparrow$0.6\%) & 0.450 ($\uparrow$0.4\%) & 0.490 ($\uparrow$0.6\%) \\
\bottomrule
\end{tabular}
\tabmargin
\vspace{-5mm}
\label{tab:overall}
\end{table*}

Table~\ref{tab:efficiency_python} shows the experiment results of time efficiency comparison in the recall step of different approaches with different sizes. Since the experiment results on different datasets are very similar, we only show the experiment results of recall efficiency on the Python dataset from the consideration of the paper length. To compare the recall efficiency of the previous deep hashing approaches with and without \tool, only the time cost in the recall step is recorded. 

First of all, we can find that \tool can reduce more than 95\% of the searching time for all sizes of the dataset and hash bits compared to previous deep Hamming distance-based hashing approaches. What's more, the efficiency of \tool is also higher than the conventional approaches, which demonstrates the effectiveness of \tool on the improvement of recall efficiency.


From Table~\ref{tab:efficiency_python}, we can also find that the time cost of the deep hashing approaches with \tool has sublinear growth while the time cost of the deep hashing approaches has superlinear growth as the size of the dataset grows, which demonstrates that the deep hashing approach with \tool is more efficient than the deep hashing approach without \tool with the larger dataset. However, we can notice that although the increasing tendency of time cost of \tool with the increase of dataset size is sublinear, it still does not meet the $O(1)$ complexity. The reason for it is that \tool contains both searching and sorting processes in the recall step. Although the time complexity of the searching process with the lookup hash tables is $O(1)$, we still need to count the appearance times of the hit candidates in each lookup hash table and sort these candidates to determine the list of recall candidates according to preset the recall number. 
It is unnecessary to worry whether the sorting process will harm the effectiveness of \tool since the previous deep hashing approaches also contain the sorting process with the time complexity of $O(n{\rm log}n)$ for the entire dataset in the recall step. Since \tool cannot recall the code candidates more than the dataset has, the upper bound of the time complexity of the sorting process in \tool is $O(n{\rm log}n)$ with the dataset containing $n$ code snippets, which is no large than the time complexity of the sorting process in previous deep hashing approaches. The efficiency of deep hashing approaches with \tool will keep increasing with the increase of the dataset compared to the previous deep hashing approaches.

\begin{tcolorbox}[width=\linewidth,boxrule=0pt,top=1pt, bottom=1pt, left=1pt,right=1pt, colback=black!15,colframe=gray!20]
In summary, \tool significantly reduces recall time compared to previous deep hashing approaches and even surpasses classical approaches like LSH. Moreover, the efficiency advantages of \tool will further increase with the expansion of the data size.
\end{tcolorbox}

\secmargin
\subsection{RQ2: What is the Effectiveness of \tool?}
\label{sec:rq2}
\secmargin

Table~\ref{tab:overall} illustrates the results of the overall performance comparison of different approaches with different code retrieval models. First of all, we can find that \tool can preserve at least 97.0\%,of the performance with all the deep hashing based code retrieval baselines in all the datasets, respectively. In addition, \tool also outperforms the conventional baselines in most metrics. These results demonstrate that \tool can retain most of the retrieval performance. 

Moreover, we can find that the performance gap between  deep hashing approaches with and without \tool shrinks when the hash codes have more bits. DASH and JDSH with \tool even outperform baselines with 256 hash bits. The reason for this performance improvement is the mechanism of \tool. 
The increase of the hash code length will directly increase the number of lookup hash tables under the setting of \tool, which can effectively increase the possibility of recall of the corresponding code. Since the hash codes are very space-efficient and the extra space cost for the increase of the hash code's length can be almost neglected. This phenomenon indicates that the problem of performance drop with \tool can be addressed by increasing hash code's length, which further demonstrates the potential of \tool.

Lastly, we can find that the performance of \tool is relatively stable under different deep hashing approaches, 
which demonstrates the generalizability of \tool. However, we can still find that there is a small performance difference under different deep hashing approaches. The reason for this phenomenon is the hashing projection distribution, which will be discussed in Section~\ref{sec:rq3}.

\begin{tcolorbox}[width=\linewidth,boxrule=0pt,top=1pt, bottom=1pt, left=1pt,right=1pt, colback=black!15,colframe=gray!20]
In summary, \tool retains more than 98\% of performance compared to previous deep hashing approaches and even outperforms some of them. Additionally, \tool significantly outperforms the classical hash table-based approach LSH when using the same number of hash tables.
\end{tcolorbox}

\secmargin
\subsection{RQ3: What is the Effectiveness of Adaptive Bits Relaxing and Iterative Training?}
\label{sec:rq3}
\secmargin

\begin{table*}[t]
\tabmargin
\footnotesize
\setlength\tabcolsep{9pt}
\centering
\revise{\caption{The comparisons among the six \tool variants with the baseline of CodeBERT.}}
\label{tab:ablation}
\resizebox{0.9\linewidth}{!}{
\begin{tabular}{lllllllllllll}
\toprule
\multirow{3}{*}{\textbf{Model}} & \multicolumn{6}{c}{\textbf{Python}} & \multicolumn{6}{c}{\textbf{Java}} \\

\cmidrule(lr){2-7} \cmidrule(lr){8-13}
 & \multicolumn{3}{c}{\textbf{128bit}} & \multicolumn{3}{c}{\textbf{256bit}} & \multicolumn{3}{c}{\textbf{128bit}} & \multicolumn{3}{c}{\textbf{256bit}}\\

\cmidrule(lr){2-4} \cmidrule(lr){5-7} \cmidrule(lr){8-10}\cmidrule(lr){11-13}
 & \textbf{R@1} & \textbf{MRR} & \textbf{N@10} & \textbf{R@1} & \textbf{MRR} & \textbf{N@10} & \textbf{R@1} & \textbf{MRR} & \textbf{N@10} & \textbf{R@1} & \textbf{MRR} & \textbf{N@10} \\
 
\midrule
CoSHC\textsubscript{NA\_NR} & 0.271 & 0.312 & 0.328 & 0.334 & 0.391 & 0.413 & 0.214 & 0.263 & 0.283 & 0.251 & 0.313 & 0.339 \\ 
CoSHC\textsubscript{A\_NR} & 0.385 & 0.460 & 0.491 & 0.411 & 0.494 & 0.529 & 0.266 & 0.337 & 0.366 & 0.291 & 0.371 & 0.403 \\ 
\hdashline
CoSHC\textsubscript{NA\_SR} & 0.417 & 0.499 & 0.533 & 0.440 & 0.533 & 0.572 & 0.301 & 0.384 & 0.418 & 0.311 & 0.400 & 0.437 \\ 
CoSHC\textsubscript{A\_SR} & 0.433 & 0.526 & 0.564 & 0.444 & 0.542 & 0.582 & 0.306 & 0.393 & 0.429 & 0.314 & 0.406 & 0.444 \\ 
\hdashline
CoSHC\textsubscript{NA\_BR} & 0.445 & 0.543 & 0.584 & 0.451 & \textbf{0.554} & \textbf{0.596} & 0.315 & 0.405 & 0.442 & \textbf{0.319} & 0.414 & 0.453 \\ 
CoSHC\textsubscript{A\_BR} & \textbf{0.447} & \textbf{0.547} & \textbf{0.589} & \textbf{0.452} & \textbf{0.554} & \textbf{0.596} & \textbf{0.316} & \textbf{0.408} & \textbf{0.445} & \textbf{0.319}  & \textbf{0.415} & \textbf{0.454}\\ 
\midrule
DJSRH\textsubscript{NA\_NR} & 0.078 & 0.086 & 0.089 & 0.124 & 0.140 & 0.146 & 0.061 & 0.072 & 0.077 & 0.110 & 0.132 & 0.141 \\ 
DJSRH\textsubscript{A\_NR} & 0.384 & 0.460 & 0.491 & 0.414 & 0.500 & 0.535 & 0.268 & 0.338 & 0.367 & 0.289 & 0.368 & 0.401\\ 
\hdashline
DJSRH\textsubscript{NA\_SR} & 0.250 & 0.289 & 0.304 & 0.319 & 0.375 & 0.397 & 0.183 & 0.225 & 0.243 & 0.249 & 0.312 & 0.339 \\ 
DJSRH\textsubscript{A\_SR} & 0.434 & 0.526 & 0.564 & 0.445 & 0.542 & 0.582 & 0.305 & 0.392 & 0.428 & 0.313 & 0.404 & 0.441 \\ 
\hdashline
DJSRH\textsubscript{NA\_BR} & 0.396 & 0.472 & 0.504 & 0.414 & 0.497 & 0.531 & 0.272 & 0.344 & 0.374 & 0.297 & 0.379 & 0.413 \\ 
DJSRH\textsubscript{A\_BR} & \textbf{0.446} & \textbf{0.546} & \textbf{0.587} & \textbf{0.451} & \textbf{0.553} & \textbf{0.596} & \textbf{0.316} & \textbf{0.409} & \textbf{0.446} & \textbf{0.319} & \textbf{0.414} & \textbf{0.452} \\ 
\midrule
DSAH\textsubscript{NA\_NR} & 0.313 & 0.365 & 0.386 & 0.375 & 0.445 & 0.474 & 0.232 & 0.288 & 0.311 & 0.270 & 0.341 & 0.370 \\ 
DSAH\textsubscript{A\_NR} & 0.388 & 0.466 & 0.498 & 0.417 & 0.502 & 0.537 & 0.268 & 0.339 & 0.369 & 0.291 & 0.371 & 0.403 \\ 
\hdashline
DSAH\textsubscript{NA\_SR} & 0.421 & 0.509 & 0.544 & 0.438 & 0.533 & 0.572 & 0.299 & 0.383 & 0.417 & 0.310 & 0.398 & 0.434 \\ 
DSAH\textsubscript{A\_SR} & 0.436 & 0.529 & 0.567 & 0.443 & 0.540 & 0.580 & 0.305 & 0.393 & 0.428 & 0.314 & 0.405 & 0.442 \\ 
\hdashline
DSAH\textsubscript{NA\_BR} & 0.441 & 0.537 & 0.577 & 0.449 & 0.550 & 0.590 & 0.312 & 0.402 & 0.438 & 0.317 & 0.410 & 0.448 \\ 
DSAH\textsubscript{A\_BR} & \textbf{0.447} & \textbf{0.547} & \textbf{0.587} & \textbf{0.452} & \textbf{0.554} & \textbf{0.596} & \textbf{0.316} & \textbf{0.409} & \textbf{0.446} & \textbf{0.319} & \textbf{0.415} & \textbf{0.454} \\ 
\midrule
JDSH\textsubscript{NA\_NR} & 0.326 & 0.384 & 0.408 & 0.384 & 0.459 & 0.489 & 0.246 & 0.307 & 0.332 & 0.281 & 0.355 & 0.385 \\ 
JDSH\textsubscript{A\_NR} & 0.388 & 0.465 & 0.497 & 0.416 & 0.502 & 0.537 & 0.269 & 0.340 & 0.369 & 0.290 & 0.370 & 0.402 \\ 
\hdashline
JDSH\textsubscript{NA\_SR} & 0.425 & 0.513 & 0.550 & 0.438 & 0.533 & 0.572 & 0.304 & 0.388 & 0.423 & 0.311 & 0.400 & 0.436 \\ 
JDSH\textsubscript{A\_SR} & 0.436 & 0.529 & 0.567 & 0.445 & 0.543 & 0.584 & 0.308 & 0.395 & 0.431 & 0.314 & 0.405 & 0.443 \\ 
\hdashline
JDSH\textsubscript{NA\_BR} & 0.441 & 0.537 & 0.577 & 0.448 & 0.549 & 0.590 & 0.314 & 0.404 & 0.441 & 0.318 & 0.411 & 0.449 \\ 
JDSH\textsubscript{A\_BR} & \textbf{0.447} & \textbf{0.547} & \textbf{0.587} & \textbf{0.452} & \textbf{0.554} & \textbf{0.596} & \textbf{0.316} & \textbf{0.409} & \textbf{0.447} & \textbf{0.319} & \textbf{0.415} & \textbf{0.452} \\ 
\bottomrule
\end{tabular}
\tabmargin
\vspace{-5mm}
}
\end{table*}

\begin{table}[t]
\vspace{-5mm}
\tabmargin
\small
\setlength\tabcolsep{8pt}
\centering
\caption{The repair ratio of adaptive bits relaxing in both code hashing model and query hashing model.}
\label{tab:fix_ratio}
\resizebox{0.8\linewidth}{!}{
\begin{tabular}{llllll}
\toprule
\multirow{2}{*}{\textbf{}}&\multirow{2}{*}{\textbf{Model}} & \multicolumn{2}{c}{\textbf{Python}} & \multicolumn{2}{c}{\textbf{Java}} \\

\cmidrule(lr){3-4} \cmidrule(lr){5-6}
& & \textbf{128bit} & \textbf{256bit} & \textbf{128bit} & \textbf{256bit}\\

\midrule
\parbox[t]{3mm}{\multirow{8}{*}{\rotatebox[origin=c]{90}{CodeBERT}}} & CoSHC\textsubscript{Code} & 0.356 & 0.358 & 0.359 & 0.370\\ 
& CoSHC\textsubscript{Query} & 0.356 & 0.358  & 0.357 & 0.369 \\ 
\cmidrule(lr){2-6}
& DJSRH\textsubscript{Code} & 0.356 & 0.377 & 0.365 & 0.387\\ 
& DJSRH\textsubscript{Query} & 0.356 & 0.378  & 0.362 & 0.384 \\ 
\cmidrule(lr){2-6}
& DSAH\textsubscript{Code} & 0.353 & 0.374 & 0.360 & 0.382\\
& DSAH\textsubscript{Query} & 0.355 & 0.374 & 0.358 & 0.382\\
\cmidrule(lr){2-6}
& JDSH\textsubscript{Code} & 0.353 & 0.374 & 0.360 & 0.385\\
& JDSH\textsubscript{Query} & 0.350 & 0.374 & 0.357 & 0.386\\
\midrule
\parbox[t]{3mm}{\multirow{8}{*}{\rotatebox[origin=c]{90}{GraphCodeBERT}}} & $\rm CoSHC_{Code}$ & 0.353 & 0.358 & 0.359 & 0.375 \\ 
& CoSHC\textsubscript{Query} & 0.355 & 0.359  & 0.356 & 0.374   \\ 
\cmidrule(lr){2-6}
 & DJSRH\textsubscript{Code} & 0.356 & 0.377 & 0.365 & 0.387\\ 
& DJSRH\textsubscript{Query} & 0.356 & 0.378  & 0.362 & 0.384 \\ 
\cmidrule(lr){2-6}
& DSAH\textsubscript{Code} & 0.350 & 0.376  & 0.357 & 0.385\\
& DSAH\textsubscript{Query} & 0.348 & 0.376 & 0.355 & 0.384\\
\cmidrule(lr){2-6}
& JDSH\textsubscript{Code} & 0.349 & 0.374 & 0.354 & 0.382\\
& JDSH\textsubscript{Query} & 0.350 & 0.374 & 0.351 & 0.381\\
\bottomrule
\end{tabular}

}
\end{table}

\begin{table}[t]
\tabmargin
\small
\centering
\setlength\tabcolsep{3pt}
\caption{Average hash bits that both code and query hashing models predicted as unknown in single hash code segment.}
\label{tab:both_zero}
\resizebox{0.8\linewidth}{!}{
\begin{tabular}{llllll}
\toprule
\multirow{2}{*}{\textbf{}}&\multirow{2}{*}{\textbf{Model}} & \multicolumn{2}{c}{\textbf{Python}} & \multicolumn{2}{c}{\textbf{Java}} \\

\cmidrule(lr){3-4} \cmidrule(lr){5-6}
& & \textbf{128bit} & \textbf{256bit} & \textbf{128bit} & \textbf{256bit}\\

\midrule
\multirow{4}{*}{CodeBERT} & $\rm CoSHC_{\tool}$ & 1.10 & 1.10 & 1.07 & 1.04 \\
& DJSRH\textsubscript{\tool} & 1.10 & 1.02 & 1.06 & 1.00 \\ 
& DSAH\textsubscript{\tool} & 1.11 & 1.04  & 1.07 & 1.01\\
& JDSH\textsubscript{\tool} & 1.11 & 1.03 & 1.07  & 0.99\\
\midrule
\multirow{4}{*}{GraphCodeBERT} & $\rm CoSHC_{\tool} $  & 1.11 & 1.09 & 1.08 & 1.01   \\ 
& DJSRH\textsubscript{\tool}  & 1.12 & 1.01 & 1.08 & 1.01   \\ 
& DSAH\textsubscript{\tool} & 1.13 & 1.03 & 1.09 & 1.01\\
& JDSH\textsubscript{\tool} & 1.13 & 1.03 & 1.09 & 1.01\\
\bottomrule
\end{tabular}
\tabmargin
}
\end{table}

Table~\ref{tab:ablation} illustrates the performance comparison of the six variants of \tool with the baseline of CodeBERT. 
There are five types of subscripts in Table~\ref{tab:ablation}, which are $\rm NA$, $\rm A$, $\rm NR$, $\rm SR$, and $\rm BR$. $\rm NA$ represents the model only splits the long hash code into segmented hash codes without iterative training. $\rm A$ represents the model splits the long hash code into several segmented hash codes with the iterative training. $\rm NR$ represents the model doesn't adopt the adaptive bits relaxing strategy. $\rm SR$ represents the model only adopts the adaptive bits relaxing strategy on the code hash model. $\rm BR$ represents the model adopts the adaptive bits relaxing strategy on both hash models. The strategy of $\rm NA$ or $\rm A$ can be combined with the strategy of $\rm NR$, $\rm SR$, or $\rm BR$ arbitrary. 

As shown in Table~\ref{tab:ablation}, $\rm Model_{A\_BR}$ achieves the best performance, which demonstrates the effectiveness of the combination of iterative strategy and adaptive bits relaxing strategy. Besides, we can find that either iterative strategy or adaptive bits relaxing strategy can greatly improve the model performance. By comparing the performance between $\rm Model_{NA}$ and $\rm Model_A$, we can find the performance for all the setting are improved. The performance of $\rm Model_A$ can be greatly improved when the performance of $\rm Model_{NA}$ is low. For example, we can find the performance of $\rm Model_{NA\_NR}$ is far from the performance of the baselines and the performance of $\rm Model_{A\_NR}$ improved a lot. By comparing the performance among $\rm Model_{NR}$, $\rm Model_{SR}$, and $\rm Model_{BR}$, we can also get similar results as above. What's more, we can find the performance improvement brought by the adaptive bits relaxing strategy is higher than the improvement brought by the iterative training strategy. $\rm Model_{NA\_BR}$  can preserve more than 98\% performance of $\rm Model_{A\_BR}$ for, CoSHC, DSAH, and JDSH baselines. 

\wcgu{Interestingly, we observed that the performance of $\rm DJSRH_{NA}$ is notably inferior to the other models. This outcome underscores the significant variations in hash codes initialized by different depth hashing methods. Fortunately, the combination of iterative training strategy and adaptive bit relaxing strategy compensates for this initial projection deficiency. As a result, the final performance of DJSRH is only slightly worse than other deep hashing approaches. This further highlights the effectiveness and versatility of \tool.}

\begin{tcolorbox}[width=\linewidth,boxrule=0pt,top=1pt, bottom=1pt, left=1pt,right=1pt, colback=black!15,colframe=gray!20]
In summary, both the adaptive bit relaxing strategy and the iterative training strategy contribute to the performance of \tool. However, the former has a larger impact on performance compared to the latter.
\end{tcolorbox}

\secmargin
\subsection{RQ4: How Many Error Bits Have Been Fixed?}
\secmargin
Table~\ref{tab:fix_ratio} illustrates the repair ratio of the adaptive bits relaxing in both hashing models. $\rm Model_{Code}$ and $\rm Model_{Query}$ are the repair ratio of the relaxed bits in the code hashing model and query hashing model, respectively. The definition of the repair ratio is the ratio of whether the bits predicted as unknown are the misaligned bits of the binary hash codes generated from the two hashing models in the initial hashing projection training process. Noted that hash bits that are relaxed by both sides of hashing model are not counted.

As shown in Table~\ref{tab:fix_ratio}, the repair ratio of all the baselines with \tool is very close. Another finding is that the repair ratio of DJSRH is slightly higher than the other two baselines. As shown in \S~\ref{sec:rq3}, the initial hash projection of DJSRH is much worse than others, which provides more space for \tool to play its advantages. In addition, we can find the repair ratio with 256 bits higher than that with 128 bits. The reason is that the relatively minimum resolution of the hash codes increases when the hash codes get longer, making it easier for \tool to distinguish which hash bits have a higher probability to make mistakes. 

Table~\ref{tab:both_zero} illustrates the average hash bits which both hashing models predicted as unknown bits in the single hash code segment. 
It is unnecessary to predict as unknown in the same bit from both the hashing models since the two hash code segments can be matched as long as the misaligned hash bit is predicted as unknown in either the code hashing model or query hashing model. On the contrary, the prediction of uncertainty in one hash bit will also reduce the hamming distance of unrelated hash codes, which may bring false positives. Therefore, the average number of unknown predictions in both code and query hashing models is smaller, and the performance of our proposed approach will be better. As shown in Table~\ref{tab:both_zero}, all the baselines with \tool have similar performance. Similar to the condition in repair ratio, DJSRH has fewer average hash bits which both hashing models predicted as unknown. Besides, we can find the average hash bits predicted as unknown from both hashing models with 256 bits is less than the average hash bits with 128 bits. The reason for this phenomenon is similar to the condition in repair ratio.

\begin{tcolorbox}[width=\linewidth,boxrule=0pt,top=1pt, bottom=1pt, left=1pt,right=1pt, colback=black!15,colframe=gray!20]
With the increasing hash code length, the repair ratio increases and the number of hash bits relaxed by both hashing models decreases. This indicates less resource waste for \tool as the hash code length increases.
\end{tcolorbox}
\secmargin
\section{Discussion}
\secmargin
\label{sec:discussion}

In this section, we will first examine how faithful \tool is to the original deep hashing methods. Next, we will explore the impact of varying segment lengths on overall performance. Finally, we will analyze \tool's performance on one-to-many datasets, where a single query may match multiple code snippets within the database.

\secmargin
\subsection{Faithfulness of \tool}
\secmargin
\revise{The proportion of code snippets that both the original deep hashing model and \tool can recall is shown in Table II in Appendix. The experimental results indicate that \tool can recall approximately 90\% of the code snippets that the original deep hashing approaches can recall. However, \tool shows an overall performance drop of only around 2\% compared to the original deep hashing approaches. This suggests that while \tool may miss some correct candidates during the recall stage compared to the original deep hashing approaches, the disparity between these code snippets and queries is often too significant for even the re-rank models (original code retrieval models) to correctly retrieve them.}

\secmargin
\subsection{Impact of Segment Lengths}
\secmargin
\revise{We chose 16 bits for the length of hash segments to ensure the segment length is a power of 2, optimizing calculation efficiency for modern 32-bit and 64-bit processors. An 8-bit segment is too short, as it only allows for 256 different hash values, leading to excessive hash collisions even if the data is evenly distributed. We also experimented with a 32-bit hash segment but found it challenging to achieve hash collisions for positive pairs with such a long segment.}

\revise{Although there is a trade-off between hash segment length and the number of relaxed hash bits, which means we can improve the hash collision probability by increasing the number of relaxed hash bits. However, it will significantly raise storage costs. For example, the storage cost for hash segments with 3 relaxed hash bits is $2^3$ times higher than that for segments without relaxed hash bits. Although storing hash codes is much more efficient than storing high-dimensional vectors, we recommend not relaxing too many hash bits within the segment.}

\revise{We have not fully explored the influence of hash segment lengths that are not powers of 2. Therefore, selecting 16 bits as the segment length might not be the optimal solution for overall performance. However, theoretically, the calculation efficiency for lengths that are not powers of 2 could be compromised due to the principles of bit operations in modern processors. Despite this, the impact may be minimal, as the bit operations for our proposed hash code are already fast enough.}

\secmargin
\subsection{Applicability to One-to-Many Dataset}
\secmargin
\revise{To make our evaluation more representative of real-world code retrieval scenarios, we also evaluate SECRET on a one-to-many dataset CoSQA+, where multiple correct answers exist for a given query. Due to the similar efficiency of hashing calculations and space limitations, we present only the overall performance comparison for this dataset. Detailed experiments can be found in Appendix Section III and Table III. The trend of SECRET’s relative performance drop on this dataset is similar to that on the one-to-one dataset, with the relative performance drop being even smaller. This further demonstrates the effectiveness of SECRET.}
\secmargin
\section{Related Work}
\label{sec:background}
\secmargin
\subsection{Code Search}
\secmargin
We briefly introduce recent deep learning-based code search approaches. NCS~\cite{SachdevLLKS018} firstly adopts FastText~\cite{BojanowskiGJM17} to embed both queries and source code into representation vectors. CRaDLe~\cite{GuLGWZXL21} further considers the dependency feature within the code snippets and adopts long short-term memory (LSTM) networks to model program dependency graph from source code. CodeBERT~\cite{FengGTDFGS0LJZ20} is a bimodal pre-trained model for programming language and natural language. GraphCodeBERT~\cite{GuoRLFT0ZDSFTDC21} considers the data flow of code during pre-training. CodeT5~\cite{0034WJH21} is a unified pre-trained encoder-decoder model trained with the identifier-aware pre-training task. Similarly, SPT-Code~\cite{NiuL0GH022} is a sequence-to-sequence pre-trained model that learns knowledge of source code, the corresponding code structure, and natural language descriptions of code. SyncoBERT~\cite{wang2021syncobert} is another pre-trained model with the objectives of Identifier Prediction and AST Edge Prediction. UniXcoder~\cite{GuoLDW0022} is a unified cross-modal pre-trained model which use mask attention matrices with prefix adapters to control the behavior of the model. Bui et al.~\cite{BuiYJ21} propose a self-supervised contrastive learning framework named Corder that  distinguishes similar and dissimilar code snippets during the training process. \wcgu{Shi et al.~\cite{ShiWGDZHZS23} propose to dynamically mask tokens to generate positive code examples for contrastive learning. Liu et al.~\cite{LiuXSMML23} propose to construct graphs and jointly learn the high-level semantics between code and queries.}

\secmargin
\subsection{Hashing Techniques}
\secmargin
In this subsection, we briefly introduce some representative hash table-based hashing approaches and Hamming distance-based cross-modal hashing approaches, which can be classified into supervised and unsupervised cross-modal hashing approaches. Besides, we also introduce the related works about the ternary hashing and segmented hashing.

\subsubsection{Hash table based approaches}
Locality Sensitive Hashing (LSH)~\cite{DatarIIM04} is one of the most popular approaches for recalling data. LSH maps high dimensional data to hash value by using random hash functions. There are several variants of LSH~\cite{BawaCG05,GanFFN12,HuangFZFN15}. Most of them need to build many lookup hash tables to guarantee the recall rate of data. Compared to these approaches, our proposed approaches can achieve similar performance with higher time efficiency and lower storage cost. Semantic Hashing~\cite{SalakhutdinovH09} is a learning-based approach and can also construct the lookup hash table for the data recall. However, most of the above approaches are single-modal approaches and do not consider the cross-modal problem.

\subsubsection{Supervised cross-modal hashing approaches} 
Bronstein et al.~\cite{BronsteinBMP10} consider the embedding of the input data from two arbitrary spaces into the Hamming space as a binary classification problem with positive and negative examples. SCM~\cite{ZhangL14} is proposed to reduce the training time complexity of most existing SMH methods. SePH~\cite{LinDH015} converts semantic affinities of training data into a probability distribution and approximates it with to-be-learned hash codes. MCSCH~\cite{YeP18} sequentially generates the hash code guided by different scale features through an RNN model with the scale information.

\subsubsection{Unsupervised cross-modal hashing approaches} 
CMFH~\cite{DingGZ14} learns unified hash codes by collective matrix factorization with latent factor model. UDCMH~\cite{WuLHLDZS18} incorporates Laplacian constraints into the objective function to preserve neighbors information. DJSRH~\cite{SuZZ19} proposes to train the hashing model with a joint-semantics affinity matrix that integrates the original neighborhood information from different modalities. Yang et al.~\cite{YangWZZL020} propose DSAH with a semantic-alignment loss function to align the similarities between features. JDSH~\cite{LiuQGZY20} utilizes Distribution-based Similarity Decision and Weighting (DSDW) for unsupervised cross-modal hashing to generate hash codes.

\subsubsection{Ternary hashing and segmented hashing}
Liu et al.~\cite{abs-2103-09173} propose a novel ternary hash encoding for learning to hash methods. They adopt Kleene logic and Łukasiewicz logic to calculate the Ternary Hamming Distance (THD) for both the training and inference stage. Liong et al.~\cite{LiongLDT20} propose to split the hash layer into several segments to expand more information during the training and finally concatenate the segmented hash codes into one binary hash code in the inference stage, which is not the same as the purpose of hashing segmentation in our proposed approach.

\secmargin
\section{Threats to Validity}
\secmargin
We have identified the following threats to validity:

\emph{Dataset Size.} 
From the consideration of the experiment cost, we only select one Python dataset and one Java dataset in our evaluation. Such an amount of data size may not be sufficient to demonstrate the performance and efficiency of \tool under huge databases.

\emph{Baseline Model Selection.} From the consideration of experiment cost, we only select two code retrieval models with three deep hashing baselines. However, it is possible that when applying \tool to other models, there is no significant time boost or the accuracy may be well preserved.

\emph{Evaluation Metrics.}
We only evaluate the proposed approach with the metric of R@1 and MRR in the overall performance experiment. However, these two metrics may not sufficiently reveal the performance gap between \tool and deep hashing baselines.
\secmargin
\section{Conclusion}
\label{sec:conclusion}
\secmargin
In this paper, we have explored the efficiency aspect of code retrieval, which has received little attention in the existing literature but is very important to the industry. Our contribution lies in a novel hashing approach built upon existing deep hashing methods. This approach involves converting long hash codes from these methods into several segmented hash codes, which are then used to construct hash tables for code candidate recall. By adopting our approach, we significantly reduce the time complexity of previous deep hashing-based approaches during the code candidates recall stage. Our experimental results demonstrate that \tool not only greatly reduces retrieval time but also achieves comparable or even higher performance than previous deep hashing approaches. 


\secmargin
\section{Acknowledgement}
\secmargin

Wenchao Gu’s and Michael R. Lyu’s work described in this paper was in part supported by the Research Grants Council of the Hong Kong Special Administrative Region, China (No. CUHK 14206921 of the General Research Fund).

\bibliographystyle{IEEEtran}
\bibliography{ref}

\end{document}


\title{Appendix}
\maketitle

\section{Faithfulness of \tool}
Table~\ref{tab:overall} presents a comparison of the overall performance of various deep hashing approaches using CodeBERT on the Python dataset with 512 hash bits. The results indicate that the performance drop is significantly reduced with 512 hash bits. This supports our claim that increasing the length of the hash codes can mitigate the performance decline.

\begin{table}[!ht]
\footnotesize
\centering
\setlength\tabcolsep{3pt}
\caption{Results of overall performance comparison of different deep hashing approaches on Python dataset (512 bits).}
\begin{tabular}{lllllllllll}
\toprule
& \textbf{Model} & \textbf{R@1} & \textbf{MRR} & \textbf{N@10} \\
 
\midrule
\parbox[t]{4mm}{\multirow{10}{*}{\rotatebox[origin=c]{90}{CodeBERT}}} & Original & 0.455 & 0.562 & 0.606 \\ 
\cmidrule(lr){2-10}
& CoSHC & 0.455 & 0.562 & 0.606 \\ 
& CoSHC\textsubscript{\tool} & 0.453 ($\downarrow$0.4\%) & 0.558 ($\downarrow$0.7\%) & 0.601 ($\downarrow$0.8\%) \\ 
\cmidrule(lr){2-10}
 & DJSRH & 0.454 & 0.561 & 0.606 \\ 
& DJSRH\textsubscript{\tool} & 0.453 ($\downarrow$0.2\%) & 0.557 ($\downarrow$0.7\%) & 0.600 ($\downarrow$1.0\%) \\ 
\cmidrule(lr){2-10}
& DSAH & 0.452 & 0.555 & 0.597 \\
& DSAH\textsubscript{\tool} & 0.454 ($\uparrow$0.4\%) & 0.558 ($\uparrow$0.5\%) & 0.600 ($\uparrow$0.5\%) \\
\cmidrule(lr){2-10}
& JDSH & 0.450 & 0.552 & 0.594 \\
& JDSH\textsubscript{\tool} & 0.454 ($\uparrow$0.9\%) & 0.559 ($\uparrow$1.2\%) & 0.602 ($\uparrow$1.3\%) \\
\midrule
\parbox[t]{4mm}{\multirow{10}{*}{\rotatebox[origin=c]{90}{GraphCodeBERT}}} & Original & 0.489 & 0.598 & 0.641 \\ 
\cmidrule(lr){2-10}
& CoSHC & 0.489 & 0.598 & 0.641 \\ 
& CoSHC\textsubscript{\tool} & 0.486 ($\downarrow$0.4\%) & 0.592 ($\downarrow$0.7\%) & 0.633 ($\downarrow$0.8\%) \\ 
\cmidrule(lr){2-10}
 & DJSRH & 0.489 & 0.598 & 0.641 \\ 
& DJSRH\textsubscript{\tool} & 0.486 ($\downarrow$0.6\%) & 0.591 ($\downarrow$1.1\%) & 0.631 ($\downarrow$1.6\%) \\ 
\cmidrule(lr){2-10}
& DSAH & 0.485 & 0.591 & 0.633 \\
& DSAH\textsubscript{\tool} & 0.487 ($\uparrow$0.4\%) & 0.592 ($\uparrow$0.2\%) & 0.634 ($\uparrow$0.2\%) \\
\cmidrule(lr){2-10}
& JDSH & 0.484 & 0.589 & 0.631 \\
& JDSH\textsubscript{\tool} & 0.486 ($\uparrow$0.4\%) & 0.591 ($\uparrow$0.3\%) & 0.633 ($\uparrow$0.3\%) \\
\bottomrule
\end{tabular}
\label{tab:overall}
\end{table}

\section{Impact of Segment Lengths}
Table~\ref{tab:faith} presents the proportion of code snippets that both the original deep hashing model and \tool can recall.

\begin{table}[!ht]
\small
\centering
\setlength\tabcolsep{6pt}
\caption{Results about proportion of code snippets that both the original deep hashing model and \tool can recall.}
\label{tab:faith}
\begin{tabular}{lllll}
\toprule
\multirow{2}{*}{\textbf{Model}} & \multicolumn{2}{c}{\textbf{Python}} & \multicolumn{2}{c}{\textbf{Java}} \\

\cmidrule(lr){2-3} \cmidrule(lr){4-5}
& \textbf{128bit} & \textbf{256bit} & \textbf{128bit} & \textbf{256bit} \\

\midrule
CoSHC\textsubscript{\tool} & 0.906 & 0.924 & 0.887 & 0.920 \\
DJSRH\textsubscript{\tool} & 0.902 & 0.915 & 0.892 & 0.905 \\
DSAH\textsubscript{\tool} & 0.922 & 0.938 & 0.908 & 0.931 \\ 
JDSH\textsubscript{\tool} & 0.926 & 0.940 & 0.909 & 0.929 \\ 
\bottomrule
\end{tabular}
\end{table}

\section{Applicability to One-to-Many Dataset}
CosQA+ includes 20,604 queries and 51,516 code snippets, with each query linked to multiple snippets. Given the dataset's limited size for model training, we adopted a zero-shot approach. Specifically, we trained the model on a different dataset intended for our primary evaluation and then applied it directly to CosQA+. The performance of \tool on this dataset is presented in Table~\ref{tab:multi}.

\begin{table}[!ht]
\footnotesize
\centering
\setlength\tabcolsep{4pt}
\caption{Results of overall performance comparison of different deep hashing approaches with different code retrieval models on CosQA+.}
\begin{tabular}{llllllll}
\toprule

\multirow{3}{*}{\textbf{Model}} & & \multicolumn{3}{c}{\textbf{128bit}} & \multicolumn{3}{c}{\textbf{256bit}} \\

\cmidrule(lr){3-5} \cmidrule(lr){6-8}
& & \textbf{R@1} & \textbf{MRR} & \textbf{N@10} & \textbf{R@1} & \textbf{MRR} & \textbf{N@10}  \\
 
\midrule
\parbox[t]{4mm}{\multirow{10}{*}{\rotatebox[origin=c]{90}{CodeBERT}}} & Original & 0.341 & 0.792 & 0.311 & 0.341 & 0.792 & 0.311 \\ 
& LSH & 0.322 & 0.678 & 0.273 & 0.337 & 0.748 & 0.299 \\ 
& BM25 & 0.337 & 0.770 & 0.297 & 0.337 & 0.770 & 0.297 \\
& TF-IDF & 0.316 & 0.654 & 0.264 & 0.337 & 0.770 & 0.264 \\
\cmidrule(lr){2-8}
& CoSHC & 0.341 & 0.783 & 0.309 & 0.341 & 0.788 & 0.310 \\ 
& CoSHC\textsubscript{\tool}  & 0.338  & 0.768 & 0.305 & 0.340 & 0.777 & 0.308 \\ 
\cmidrule(lr){2-8}
 & DJSRH & 0.339 & 0.775 & 0.307 & 0.341 & 0.786 & 0.310 \\ 
& DJSRH\textsubscript{\tool} & 0.338 & 0.766 & 0.304 & 0.340 & 0.777 & 0.307 \\ 
\cmidrule(lr){2-8}
& DSAH & 0.337 & 0.756 & 0.301 & 0.338 & 0.762 & 0.303 \\
& DSAH\textsubscript{\tool} & 0.337 & 0.764 & 0.304 & 0.340 & 0.776 & 0.308 \\
\cmidrule(lr){2-8}
& JDSH & 0.334 & 0.748 & 0.298 & 0.336 & 0.756 & 0.301 \\
& JDSH\textsubscript{\tool} & 0.338 & 0.766 & 0.304 & 0.340 & 0.777 & 0.308 \\
\midrule
\parbox[t]{4mm}{\multirow{10}{*}{\rotatebox[origin=c]{90}{GraphCodeBERT}}} & Original & 0.355 & 0.818 & 0.321 & 0.355 & 0.818 & 0.321 \\ 
& LSH & 0.329 & 0.688 & 0.278 & 0.348 & 0.767 & 0.307 \\ 
& BM25 & 0.351 & 0.790 & 0.310  & 0.351 & 0.790 & 0.310  \\
& TF-IDF & 0.323 & 0.666 & 0.268 & 0.323 & 0.666 & 0.268 \\
\cmidrule(lr){2-8}
 & CoSHC & 0.354 & 0.809 & 0.320 & 0.355 & 0.814 & 0.321 \\ 
& CoSHC\textsubscript{\tool} & 0.351 & 0.789 & 0.314 & 0.353 & 0.800 & 0.317 \\ 
\cmidrule(lr){2-8}
& DJSRH & 0.353 & 0.803 & 0.318 & 0.355 & 0.812 & 0.321 \\ 
& DJSRH\textsubscript{\tool} & 0.351 & 0.790 & 0.314 & 0.353 & 0.802 & 0.318  \\ 
\cmidrule(lr){2-8}
& DSAH & 0.349 & 0.779 & 0.311 & 0.350 & 0.787 & 0.313 \\
& DSAH\textsubscript{\tool} & 0.351 & 0.789 & 0.314 & 0.352 & 0.799 & 0.317 \\
\cmidrule(lr){2-8}
& JDSH & 0.348 & 0.776 & 0.309 & 0.349 & 0.782 & 0.312 \\
& JDSH\textsubscript{\tool} & 0.350 & 0.790 & 0.314 & 0.352 & 0.801 & 0.317 \\
\bottomrule
\end{tabular}
\label{tab:multi}
\end{table}
